\documentclass[aps,pre,twocolumn,showpacs,floatfix]{revtex4}
\usepackage{bm}
\usepackage{graphicx}

\begin{document}
\DeclareGraphicsExtensions{.jpg,.mps,.png,.tif,.eps}
\title{Localized and stationary light wave modes in dispersive media}
\author{Miguel A. Porras}
\affiliation{Departamento de F\'{\i}sica Aplicada, ETSIM, Universidad Polit\'ecnica de Madrid, Rios
Rosas 21, 28003 Madrid, Spain}
\author{Paolo Di Trapani}
\affiliation{INFM and Department of Chemical, Physical and Mathematical Sciences, University of
Insubria, Via Valleggio 11, 22100 Como, Italy}

\begin{abstract}
In recent experiments, localized and stationary pulses have been generated in second-order nonlinear
processes with femtosecond pulses, whose asymptotic features relate with those of nondiffracting and
nondispersing polychromatic Bessel beams in linear dispersive media. We investigate on the nature of
these linear waves, and show that they can be identified with the X-shaped (O-shaped) modes of the
hyperbolic (elliptic) wave equation in media with normal (anomalous) dispersion. Depending on the
relative strengths of mode phase mismatch, group velocity mismatch with respect to a plane pulse,
and of the defeated group velocity dispersion, these modes can adopt the form of pulsed Bessel
beams, focus wave modes, and X-waves (O-waves), respectively.
\end{abstract}


\maketitle

\section{Introduction}

Stationary, temporally and spatially localized, X-shaped optical wave packets, having a duration of
a few tens of femtoseconds and spot size of a few microns, have been recently observed to be
spontaneously generated in dispersive nonlinear materials from a standard laser wave packet
\cite{TRAPPRL2003,JEDRPRE2003,VA2001}. Balancing between second-order or Kerr nonlinearity, group
velocity dispersion (GVD), angular dispersion and diffraction, has been suggested to act as a kind
of mode-locking mechanism that drives pulse reshaping and keeps the interacting waves trapped, phase
and group-matched \cite{CONTIPRL2003,TRILLO,CONTI}.

The purpose of the present paper is to investigate on the nature of these waves. The main hypothesis
underlying our investigation is that these nonlinearly generated X-shaped waves behave
asymptotically as linear waves. This assumption is based, first, on the observed stationarity, not
only of the central hump of the wave packet, but also of its asymptotic, low-intensity, conical part
\cite{TRAPPRL2003,JEDRPRE2003,VA2001}, stationarity that cannot be attributed to nonlinear wave
interactions, but to some linear mechanism of compensation between material and angular dispersion.
Indeed, several kinds of linear polychromatic versions of Bessel beams \cite{DURNIN}, as Bessel-X
pulses \cite{SO96,SO97}, pulsed Bessel beams \cite{PO01OL,PO02OC}, subcycle Bessel-X pulses or focus
wave modes \cite{ORLOV1}, and envelope X waves \cite{POOL2003}, with the capability of maintaining
transversal and temporal (longitudinal) localization in dispersive linear media, have been described
in recent years (for an unified description and the extension to media with anomalous dispersion,
see also Ref. \onlinecite{POPRE2003}). In contrast to free-space X-waves \cite{LU} and Bessel-X
pulses \cite{SAALP,SAAPRL}, stationarity in dispersive media requires the introduction of an
appropriate amount of cone angle dispersion that leads to the cancellation of material GVD with cone
angle dispersion-induced GVD \cite{SO96,SO97,PO01OL,PO02OC,ORLOV1,POOL2003,POPRE2003}. Second,
polychromatic Bessel beams, with or without angular dispersion \cite{POPRE2003}, have the ability of
propagating at rather arbitrary effective phase and group velocities in dispersive media, as has to
be done by the phase matched and mutually trapped fundamental \cite{VA2001} and second harmonic
\cite{TRILLO} nonlinearly generated X waves.

For these reasons, in this paper we present a new and more comprehensive description of localized
and stationary optical waves in linear dispersive media, henceforth called {\em wave modes}, that is
particularly suitable for understanding and predicting the spatiotemporal features of the nonlinear
X waves generated in experiments. On the linear hand, this description allows us to predict the
existence of new kinds of wave modes, and classify all them according to the values of a few
physically meaningful parameters.

Each wave mode is specified by the values of the defeated material GVD, the mode group velocity
mismatch (GVM) and phase mismatch (PM) with respect to a plane pulse of the same carrier frequency
in the same medium. Wave modes are then shown (Section \ref{WM}) to belong to two broad categories:
hyperbolic modes, with X-shaped spatiotemporal structure, if material dispersion is normal, or
elliptic modes, with O-shaped structure, if material dispersion is anomalous \cite{VALIULIS}. In
Section \ref{CLASS} we show that each wave mode can adopt the approximate form of 1) a pulsed Bessel
beam (PBB), 2) an {\em envelope focus wave mode} (eFWM), or 3) an envelope X (eX) wave in normally
dispersive media [{\em envelope O} (eO) {\em wave} in anomalously dispersive media], according that
the mode bandwidth makes PM, GVM or defeated GVD, respectively, to be dominant mode characteristic
on propagation. This classification allows us to understand the spatiotemporal features of wave
modes in dispersive media in terms of a few parameters (the characteristic PM, GVM and GVD lengths),
including modes with mixed pulsed Bessel, focus wave mode, and X-like (O-like) structure.

The above description is obtained from the paraxial approximation to wave propagation. We choose
this approach because of its wider use in nonlinear optics, and because it leads to simpler
expressions in terms of parameters directly linked to the physically relevant properties of the mode
and dispersive medium. In Section \ref{NP} we compare the paraxial and the more exact nonparaxial
approaches, to show that the paraxial approach is accurate enough for the description of wave modes
currently generated by linear optical devices \cite{SO97,PO01OL} and in nonlinear wave mixing
processes \cite{TRAPPRL2003,JEDRPRE2003,VA2001}.

\section{Wave modes of the paraxial wave equation} \label{WM}

We start by considering the propagation of a three-dimensional wave packet $E(\bm x_\perp,z,t)=A(\bm
x_\perp,z,t)\exp(-i\omega_0 t+ik_0z)$  [$\bm x_\perp\equiv (x,y)$] of a certain optical carrier
frequency $\omega_0$, subject to the effects of diffraction and dispersion of the material medium.
Within the paraxial approximation, and up to second order in dispersion, the propagation of
narrow-band pulses is ruled by the equation
\begin{equation}\label{PWE}
\partial_z A =\frac{i}{2k_0}\Delta_\perp A- i\frac{k_0^{\prime\prime}}{2}
\partial_\tau^2 A ,
\end{equation}
where $z$ is the propagation direction, $\tau= t-k'_0z$ is the local time, $\Delta_\perp\equiv
\partial_x^2+\partial_y^2$, and $k^{(i)}_0\equiv
\partial_\omega^{(i)}k(\omega)|_{\omega_0}$, with $k(\omega)$ the propagation
constant in the medium.  Eq. (\ref{PWE}) is valid for a narrow envelope spectrum $\hat A(\bm
x_\perp,z,\Omega)$ around $\Omega\equiv\omega-\omega_0=0$, that is, for bandwidths
\begin{equation}\label{COND1}
\Delta\Omega\ll \omega_0 ,
\end{equation}
a condition that requires at least few carrier oscillations to fall within the envelope $A$.

We search for stationary and localized solutions of Eq. (\ref{PWE}) in the wide sense that the {\em
intensity} does not depend on $z$ in a reference frame moving at {\em some velocity}. These
solutions must then be of the form
\begin{equation}\label{ANSATZ}
A(x,y,\tau,z)=\Phi(x,y,\tau+\alpha z) \exp(-i\beta z) .
\end{equation}
The free parameters $\alpha$ and $\beta$ are assumed to be small in the sense that
\begin{eqnarray}
|\alpha|&\ll& k'_0, \label{COND2}\\
|\beta|&\ll& k_0, \label{COND3}
\end{eqnarray}
so that the group velocity $1/(k'_0-\alpha)$ and phase velocity $\omega_0/(k_0-\beta)$ of the wave
differ slightly from those of a plane pulse of the same carrier frequency in the same material,
$1/k'_0$ and $\omega_0/k_0$, respectively.

\begin{figure}[h!]
\begin{center}
\includegraphics[angle=90,width=4.2cm]{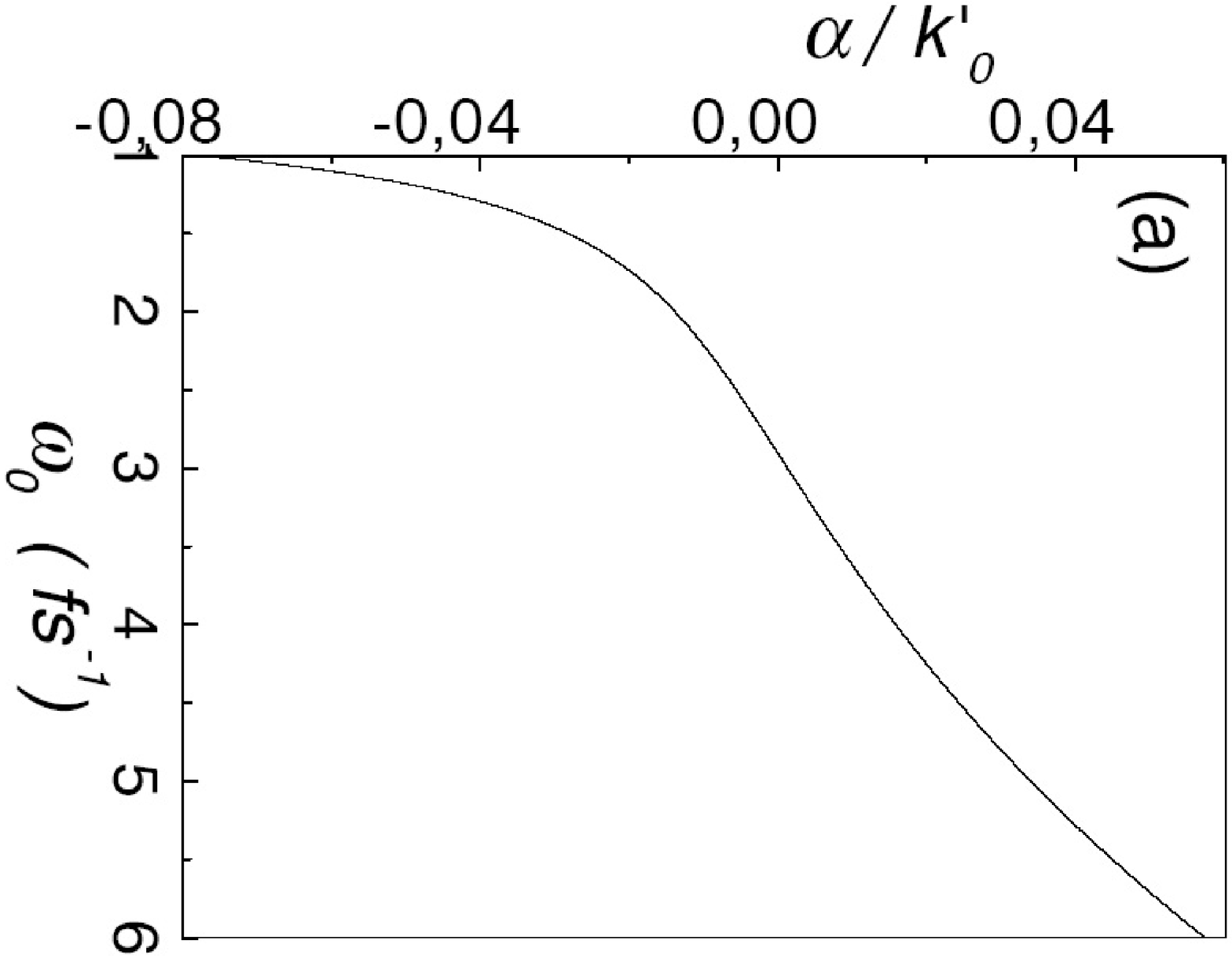}
\includegraphics[angle=90,width=4.2cm]{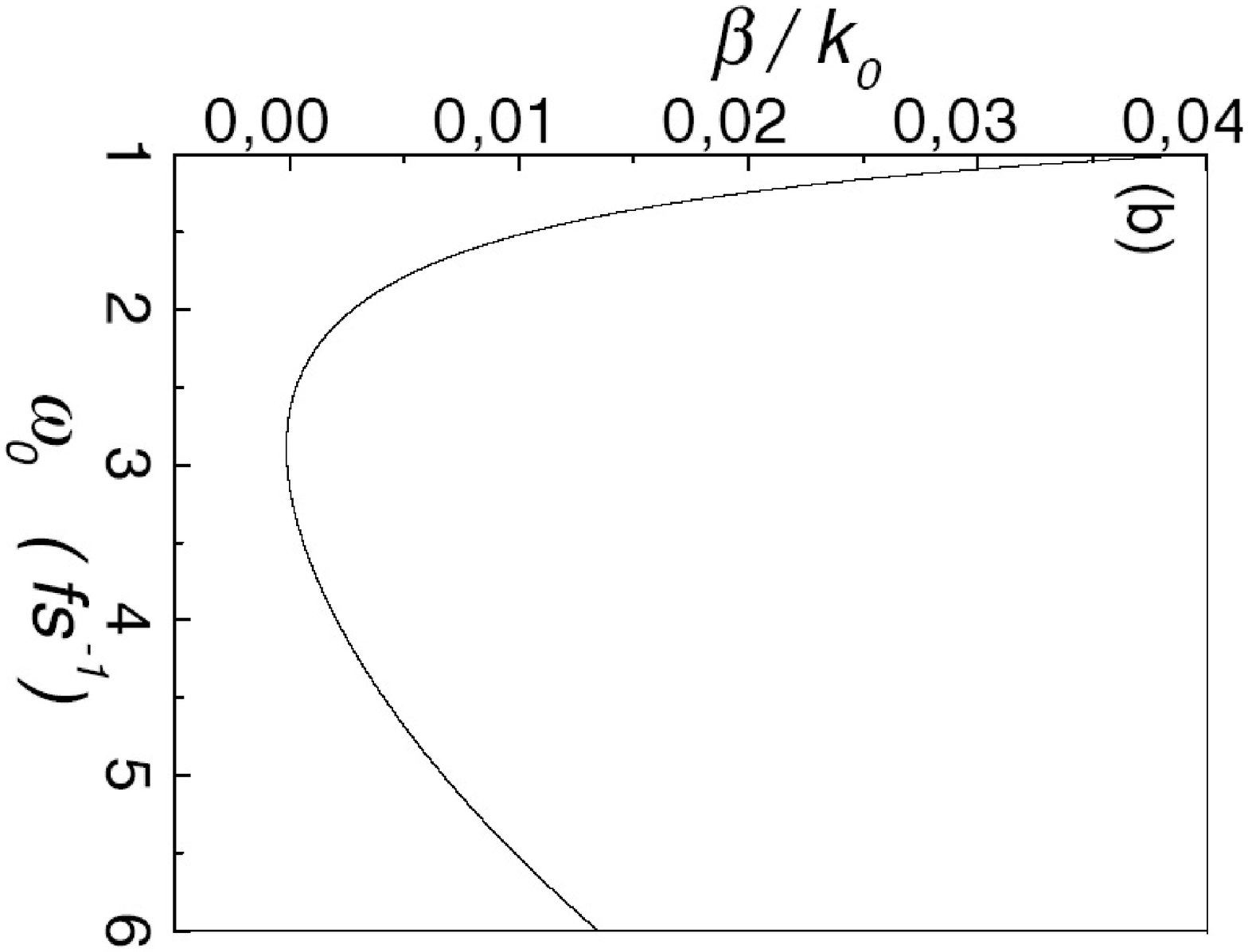}
\end{center}
\caption{\label{newfig1} Values of $\alpha$ and $\beta$ of the localized and stationary,
second-harmonic waves of different carrier frequencies $\omega_0$ for phase and group matching with
the fundamental wave in the process of oo-e second harmonic generation in LBO at room temperature.
Dispersion formulas for the refraction index are taken from Ref. \cite{HANDBOOK}}.
\end{figure}

Under the assumption of asymptotic linear behavior of nonlinear X-waves, we can get some insight on
the possible values of $\alpha$ and $\beta$ of nonlinear X-waves on the only basis of the linear
dispersive properties of the medium. If, for instance, a pulse of frequency $\omega_F$ generates a
stationary and localized second harmonic pulse ($\omega_0=2\omega_F$) travelling at the same group
and phase velocities as the fundamental pulse \cite{TRILLO}, we must have $k'_0-\alpha = k'_F$ and
$k_0-\beta=2k_F$, that is, $\alpha=k'_F-k'_0$ and $\beta=\Delta k\equiv k_0-2k_F$. For illustration,
Fig. \ref{newfig1} shows the values of $\alpha$ and $\beta$ of the second harmonic pulse in lithium
triborate (LBO) as a function of its carrier frequency $\omega_0$. Note also that $|\alpha|$ and
$|\beta|$ satisfy the conditions (\ref{COND2}) and (\ref{COND3}) for any carrier frequency in the
entire visible range and beyond.

In Section \ref{NP}, a nonparaxial approach to the problem stated above will be performed. It will
be shown that the paraxial and nonparaxial descriptions yield substantially the same results if
conditions (\ref{COND1}), (\ref{COND2}) and (\ref{COND3}) are satisfied, as is the case of the
experiments and numerical simulations demonstrating the spontaneous generation of X-type waves
\cite{TRAPPRL2003,JEDRPRE2003,VA2001}.

Equation (\ref{PWE}) with ansatz (\ref{ANSATZ}) yields
\begin{equation}\label{PWE2}
\Delta_\perp \Phi -k_0k_0^{\prime\prime}\partial_\tau^2\Phi+ 2ik_0\alpha\partial_\tau \Phi +2k_0
\beta \Phi =0,
\end{equation}
for the reduced envelope $\Phi$, or the Helmholtz-type equation $\Delta_\perp \hat \Phi
+K^2(\Omega)\hat\Phi =0$, for its temporal spectrum $\hat \Phi(x,y,\Omega)$, where
\begin{equation}\label{DISP1}
K(\Omega) = \sqrt{2k_0\left(\beta + \alpha\Omega + \frac{1}{2}k_0^{\prime\prime}\Omega^2 \right)}
\end{equation}
will be referred to as the (transversal) dispersion relation since it relates the modulus $K$ of the
transversal component of the wave vector with the detuning $\Omega$ of each monochromatic wave
component from the carrier frequency $\omega_0$. For $\Omega$ such that $K(\Omega)$ is real, the
Helmholtz equation admits the bounded, cylindrically symmetric, Bessel-type solution $\hat \Phi
(r,\Omega) =\hat f(\Omega) J_0[K(\Omega) r]$, where $\hat f(\Omega)$ is an arbitrary spectral
amplitude and $J_0(\cdot)$ the Bessel function of zero order and first class \cite{GRA}. By inverse
Fourier transform we can write the expression
\begin{eqnarray}
\Phi_{\alpha,\beta}(r,\tau+\alpha z) &=& \frac{1}{2\pi}\int_{\mbox{$K(\Omega)$ real}}
            d\Omega \hat f(\Omega)  \nonumber \\
   &\times & J_0[K(\Omega)r] \exp[- i\Omega(\tau+\alpha z)]   \label{MODESI}
\end{eqnarray}
for the reduced envelope of the cylindrically symmetric wave modes, or localized, propagation
invariant solutions of the paraxial wave equation, in the sense explained above. As indicated, the
integration domain extends over frequencies $\Omega$ such that the dispersion curve $K(\Omega)$ is
real. According to Eq. (\ref{MODESI}), a wave mode $\Phi_{\alpha,\beta}$ is composed of locked
monochromatic Bessel beams whose frequencies and radial wave vectors are linked by a specific
dispersion relation $K(\Omega)$, and whose relative weights are determined by a certain spectral
amplitude $\hat f(\Omega)$.

\begin{figure}[h!]
\begin{center}
\includegraphics[width=6.5cm]{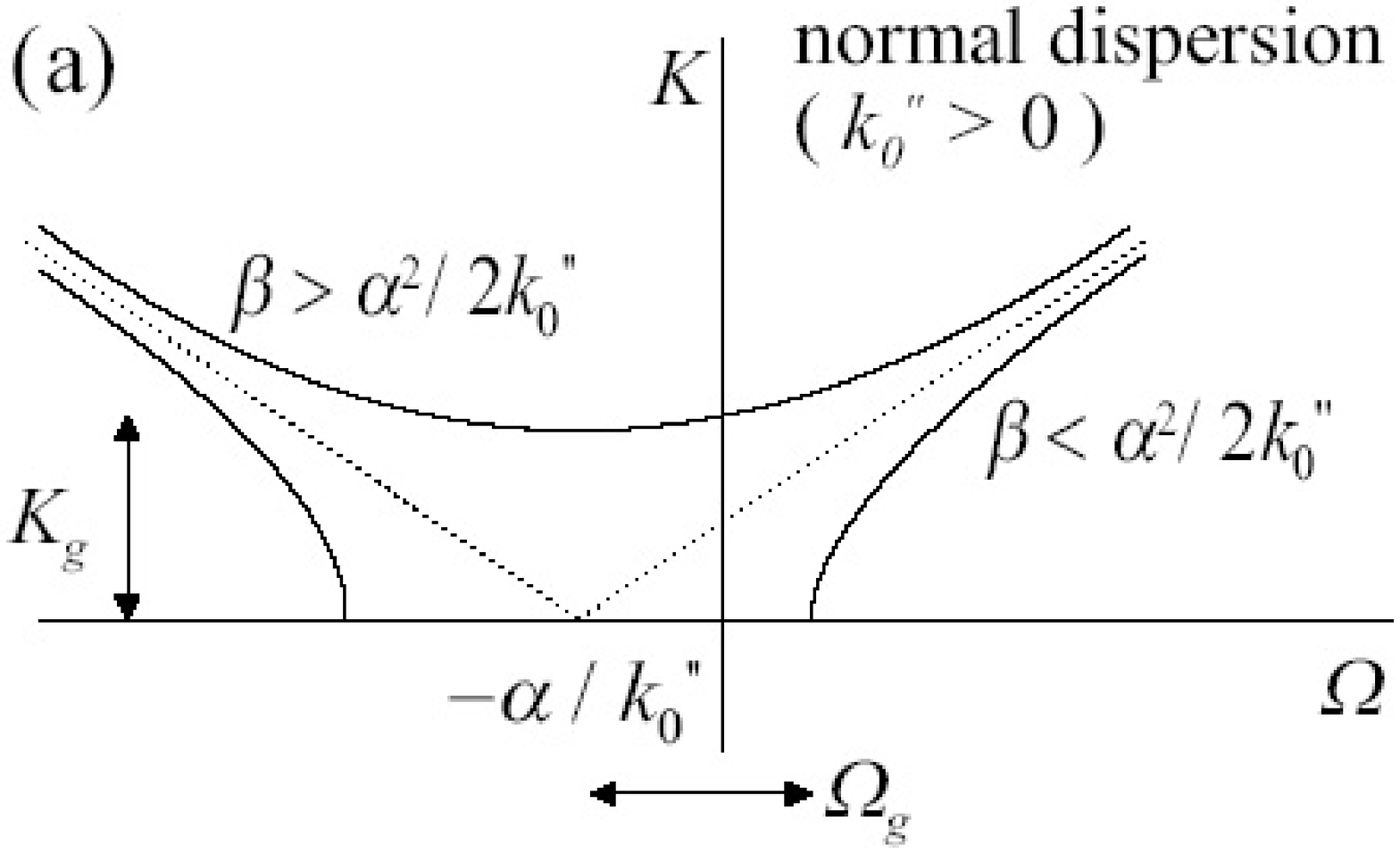}

\includegraphics[width=6.5cm]{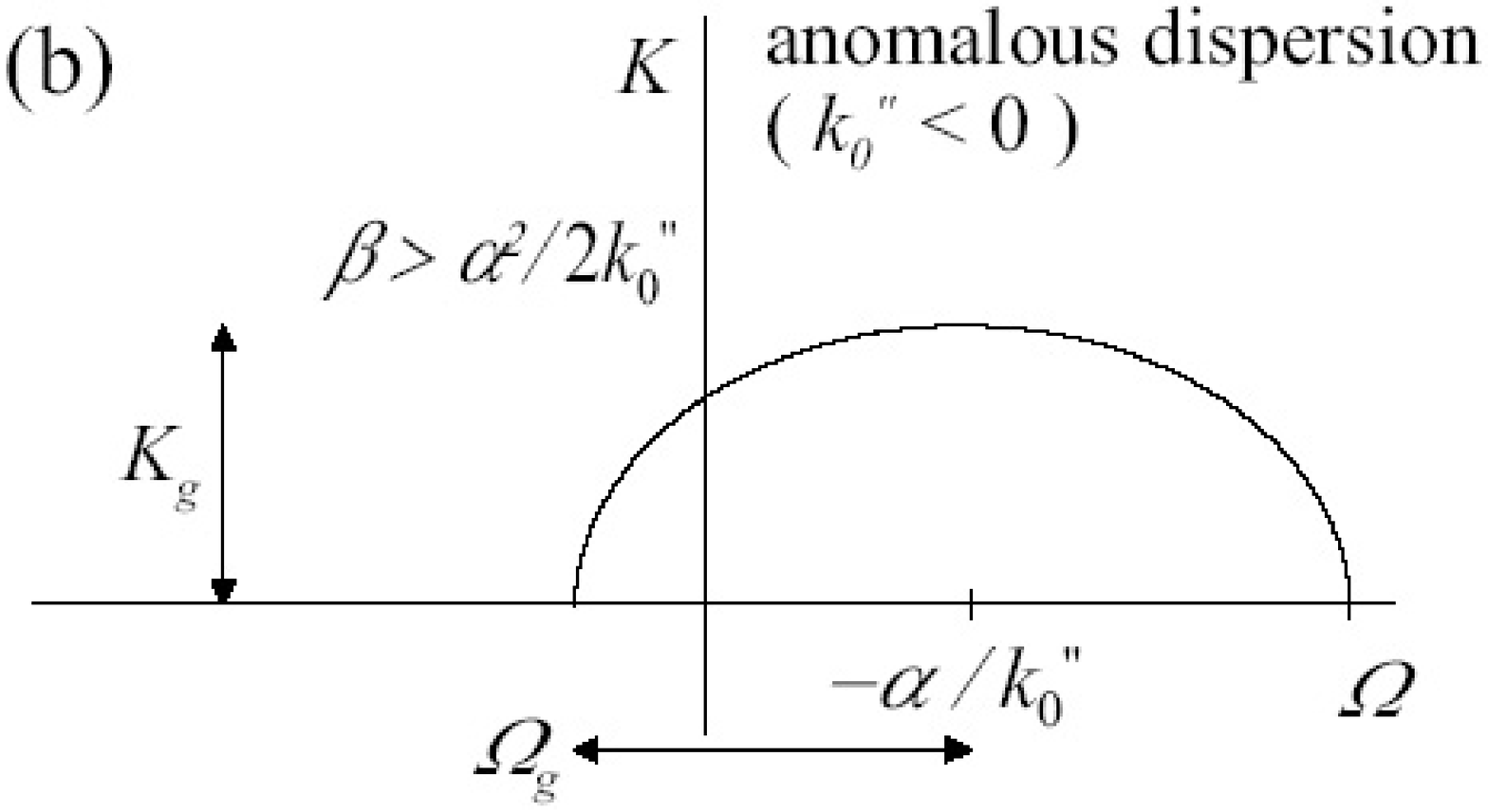}
\end{center}
\caption{\label{fig1} Dispersion curve of the wave modes in a medium with (a) mormal dispersion, (b)
anomalous dispersion.}
\end{figure}

As shown in Figs. \ref{fig1}(a) and (b), the form of the dispersion curve $K(\Omega)$ reflects the
underlying hyperbolic or elliptic geometries of the paraxial wave equation (\ref{PWE}) in the
respective cases of propagation in media with normal or anomalous dispersion. For normal dispersion
($k_0^{\prime\prime}>0$), $K(\Omega)$ is in fact a single-branch vertical hyperbola if
$\beta>\alpha^2/2k_0^{\prime\prime}$, and a two-branch horizontal hyperbola if
$\beta<\alpha^2/2k_0^{\prime\prime}$ [see Fig. \ref{fig1}(a)]. For anomalous dispersion
($k_0^{\prime\prime}<0$), $K(\Omega)$ takes real values only if
$\beta>\alpha^2/2k_0^{\prime\prime}$, in which case the dispersion curve is an ellipse [see Fig.
\ref{fig1}b]. It is also convenient to introduce the (real or imaginary) frequency gap
\begin{equation}
\Omega_g\equiv\sqrt{\frac{\alpha^2}{k_0^{\prime\prime 2}}- \frac{2\beta}{k_0^{\prime\prime}}},
\end{equation}
and radial wavevector gap
\begin{equation}
K_g\equiv \sqrt{-k_0 k_0^{\prime\prime}\Omega_g^2}.
\end{equation}
When $\Omega_g$ and $K_g$ are real, they represent actual frequency and radial wavevector gaps in
the dispersion curve $K(\Omega)$, as illustrated in Fig. \ref{fig1}. In any case, their moduli
characterize the scales of variation of the frequency and radial wavevector in the dispersion
curves.

Closely connected with the dispersion curve are the so-called {\em impulse response} wave modes
$\Phi_{\alpha,\beta}^{(i)}(r,\tau+\alpha z)$, or modes with $\hat f(\Omega)=1$. As seen in Fig.
\ref{fig1}, the structure of $\Phi_{\alpha,\beta}^{(i)}$ in space and time closely resembles that of
the dispersion curve in the $K$--$\Omega$ plane, but at radial and temporal scales of variation
determined by the reciprocal quantities $|K_g|^{-1}$ and $|\Omega_g|^{-1}$, respectively. Eq.
(\ref{MODESI}) with $\hat f(\Omega)=1$ and the change $\Omega'= \Omega+\alpha/k_0^{\prime\prime}$
yields
\begin{eqnarray}
\Phi_{\alpha,\beta}^{(i)}(r,z,\tau)
&=& \frac{1}{2\pi}\int_{\mbox{$K(\Omega')$ real}} d\Omega'  \nonumber \\
               &\times& J_0\left[K(\Omega') r\right]\exp[-i\Omega'(\tau+\alpha z)]  \nonumber \\
&\times&\exp\left[i\frac{\alpha}{k_0^{\prime\prime}}(\tau+\alpha z)\right] , \label{MODESII}
\end{eqnarray}
where
\begin{equation}
K(\Omega') =\sqrt{kk_0^{\prime\prime}\left(\Omega^{\prime 2}
          +\frac{2\beta}{k_0^{\prime\prime}}-\frac{\alpha^2}{k_0^{\prime\prime
          2}}\right)} .
\end{equation}
The integral in Eq. (\ref{MODESII}) can be performed in all possible cases from formulae 6.677.3
(for $k_0^{\prime\prime}>0$, $\beta>\alpha^2/2k_0^{\prime\prime}$),6.677.2 (for
$k_0^{\prime\prime}>0$, $\beta<\alpha^2/2k_0^{\prime\prime}$) and 6.677.6 (for
$k_0^{\prime\prime}<0$, $\beta>\alpha^2/2k_0^{\prime\prime}$) of Ref. \cite{GRA}, to yield the
closed form expression for impulse response modes

\begin{widetext}

\begin{figure}
\begin{center}
\includegraphics[width=16cm]{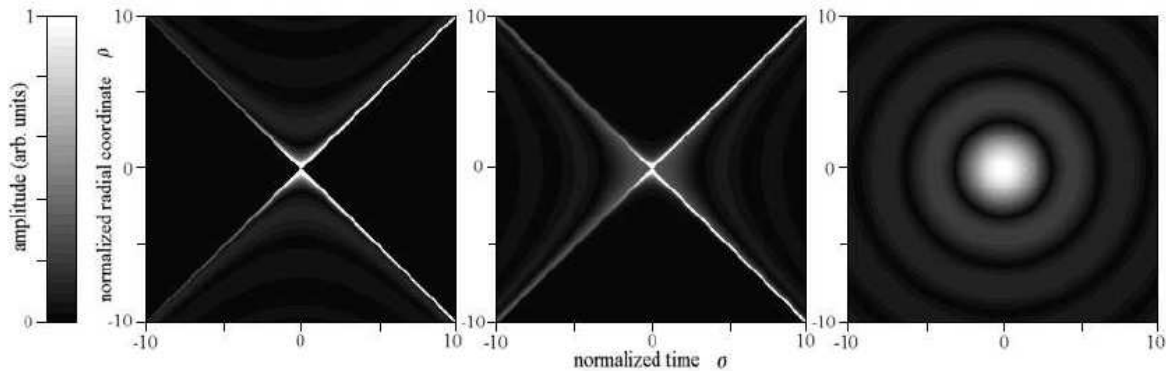}
\end{center}
\caption{\label{fig2} Gray-scale plot of the amplitude $|\Phi_{\alpha,\beta}^{(i)}|$ of the impulse
response wave modes. (a) Normal dispersion $k_0^{\prime\prime}>0$ with
$\beta>\alpha^2/2k_0^{\prime\prime}$ ($\Omega_g$ imaginary, transversal wave vector gap $K_g$ real).
(b) Normal dispersion $k_0^{\prime\prime}>0$ with $\beta<\alpha^2/2k_0^{\prime\prime}$ (detuning gap
$\Omega_g$ real, $K_g$ imaginary). (c) Anomalous dispersion $k_0^{\prime\prime}<0$ with
$\beta>\alpha^2/2k_0^{\prime\prime}$ ($\Omega_g$ and $K_g$ real). Normalized local time and radial
coordinate are defined as $\sigma=|\Omega_g|(\tau+\alpha z)$ and $\rho=|K_g|r$, respectively}
\end{figure}
\begin{eqnarray}\label{IMODESI}
\Phi^{(i)}_{\alpha,\beta}(r,\tau+\alpha z)
&=&\frac{1}{2\pi}\left[\frac{1}{\sqrt{k_0k_0^{\prime\prime}r^2-(\tau+\alpha z)^2}}
\exp\left\{i\left[\sqrt{\frac{2\beta}{k_0^{\prime\prime}}-\frac{\alpha^2}{k_0^{\prime\prime 2}}}
\sqrt{k_0k_0^{\prime\prime}r^2-(\tau+\alpha z)^2}\right]\right\}  \;\; +\;\; \mbox{C. C.}\right] \nonumber \\
     &\times&\exp\left[\frac{i\alpha}{k_0^{\prime\prime}}(\tau+\alpha z) \right]  ,
\end{eqnarray}

\end{widetext}
or, in terms of the frequency and radial wavevector gaps
\begin{equation}\label{IMODESII}
\Phi_{\alpha,\beta}^{(i)}=\frac{1}{2\pi}\left[\Omega_g\frac{\exp(iR)}{iR}+ \mbox{C.
C.}\right]\exp\left[\frac{i\alpha}{k_0^{\prime\prime}}(\tau+\alpha z)\right],
\end{equation}
where $R=[K_g^2r^2+\Omega_g^2(\tau+\alpha z)^2]^{1/2}$.

As shown in Fig. \ref{fig1}(a), for $k_0^{\prime\prime}>0$ and $\beta>\alpha^2/2k_0^{\prime\prime}$
($\Omega_g$ imaginary and $K_g$ real), the impulse response wave mode is singular in the cone
$r=|(\tau+\alpha z)|/\sqrt{k_0k_0^{\prime\prime}}$, is zero for $r<|(\tau+\alpha
z)|/\sqrt{k_0k_0^{\prime\prime}}$ (within the cone), and decays as $1/r$ for $r>|(\tau+\alpha
z)|/\sqrt{k_0k_0^{\prime\prime}}$ (out of the cone). The radial beatings in this region, of period
$2\pi/K_g$, are a consequence of the radial wave vector gap $K_g$.

Figure \ref{fig2}(b) shows the impulse response mode for $k_0^{\prime\prime}>0$ and
$\beta<\alpha^2/2k_0^{\prime\prime}$ ($\Omega_g$ real and $K_g$ imaginary). As in the previous case,
the mode is singular at the cone $r=|(\tau+\alpha z)|/\sqrt{k_0k_0^{\prime\prime}}$, but damped
oscillations are now temporal, of period $2\pi/\Omega_g$, as corresponds to the frequency gap
$\Omega_g$ in the dispersion curve. Out of the cone [$r>|(\tau+\alpha
z)|/\sqrt{k_0k_0^{\prime\prime}}$] the mode is exponentially localized.

Modes in media with anomalous dispersion, i.e., with $k_0^{\prime\prime}<0$ and
$\beta>\alpha^2/2k_0^{\prime\prime}$ (real $\Omega_g$ and $K_g$), exhibit rather different
characteristics [Fig. \ref{fig2}(c)]. These modes are no longer singular and of X-type, but regular
and, say, of O-type. The damped oscillations decay temporally and radially as $1/t$ and $1/r$,
respectively, with periods $2\pi/\Omega_g$ and $2\pi/K_g$. The absence of singularities is a
consequence of the actual limitation that the elliptic dispersion curve imposes to the uniform
spectrum $\hat f(\Omega)=1$.

\section{Classification of wave modes}\label{CLASS}

Numerical integration of Eq. (\ref{MODESI}) with a given dispersion curve (specified by the values
of $\alpha$, $\beta$ and $k_0^{\prime\prime}$) but different (bell-shaped) spectral amplitude
functions $\hat f(\Omega)$ having also different (but finite) bandwidths $\Delta\Omega$
[alternatively, numerical integration of
\begin{equation}
\Phi_{\alpha,\beta}(r,\tau+\alpha z) = \int_{-\infty}^{\infty} d\sigma
\Phi_{\alpha,\beta}^{(i)}(r,\tau+\alpha z -\sigma)
\end{equation}
where $f(\tau)$ is the inverse Fourier transform of $\hat f(\Omega)$], shows much richer and complex
spatiotemporal features in comparison with the case of infinite bandwidth. These features strongly
depend on the choice of the spectral bandwidth $\Delta\Omega$, while no essentially new properties
arise from the specific choice of $\hat f(\Omega)$ (Gaussian, Lorentzial, two-side
exponential\dots). Modes with finite bandwidth may exhibit mixed, more or less pronounced radial and
temporal oscillations, along with incipient or strong X-wave (O-wave), focus wave mode or Bessel
structure, as explained throughout this section (see also the following figures). The purpose of
this section is to perform a simple, comprehensive classification of wave modes in dispersive media.
In the remainder of this paper, $\Delta\Omega$ will refer to any suitable definition of half-width
of the bell-shaped spectral amplitude function $\hat f(\Omega)$.

Given a mode of parameters $\alpha$ and $\beta$ satisfying conditions (\ref{COND2}) and
(\ref{COND3}), propagating in a dispersive material with GVD $k_0^{\prime\prime}$, and some spectral
bandwidth $\Delta\Omega$ satisfying (\ref{COND1}), we have found it convenient to define the three
following characteristic lengths: 1) the mode PM length
\begin{equation}
L_p\equiv \frac{1}{\beta} ,
\end{equation}
2) the mode walk-off, or GVM length
\begin{equation}
L_w\equiv \frac{1}{\alpha\Delta\Omega},
\end{equation}
measuring, respectively, the axial distances at which the mode becomes phase mismatched and walks
off {\em with respect to a plane pulse of the same spectrum in the same medium}, and 3) the GVD
length
\begin{equation}
L_d\equiv \frac{1}{k_0^{\prime\prime}(\Delta \Omega)^2},
\end{equation}
or distance at which the mode (invariable) duration differs significantly from that of the
(broadening) plane pulse. Note that, as defined, $L_p$, $L_w$ and $L_d$ can be positive or negative.
In terms of the mode lengths the transversal dispersion relation (\ref{DISP1}) takes the form
\begin{equation}\label{DISP2}
K(\Omega)=\sqrt{2k_0\left(L_p^{-1}+L_w^{-1}\Omega_n+ \frac{1}{2} L_d^{-1}\Omega_n^2 \right)} ,
\end{equation}
where $\Omega_n=\Omega/\Delta\Omega$ is the normalized detuning, which ranges in $[-1,+1]$ for
$\Omega$ within the bandwidth $\Delta\Omega$. Then, they are the values of the mode lengths $L_p$,
$L_w$ and $L_d$ that determine the form of the dispersion curve within the spectral bandwidth, and
hence the parameters that determine the spatiotemporal structure of the mode, as shown throughout
this section. We analyze here three extreme cases, namely,
\begin{eqnarray}
|L_p|&\ll& |L_w|, |L_d| \,\,\,\, \mbox{PM-dominated case} \nonumber \\
|L_w|&\ll& |L_p|, |L_d| \,\,\,\, \mbox{GVM-dominated case} \nonumber \\
|L_d|&\ll& |L_p|, |L_w| \,\,\,\, \mbox{GVD-dominated case} \nonumber
\end{eqnarray}
that represent three well-defined, opposite experimental situations, and that allow us also to
understand, at least qualitatively, the features of general, intermediate cases.

\begin{figure}
\begin{center}
\includegraphics[angle=90,width=4.2cm]{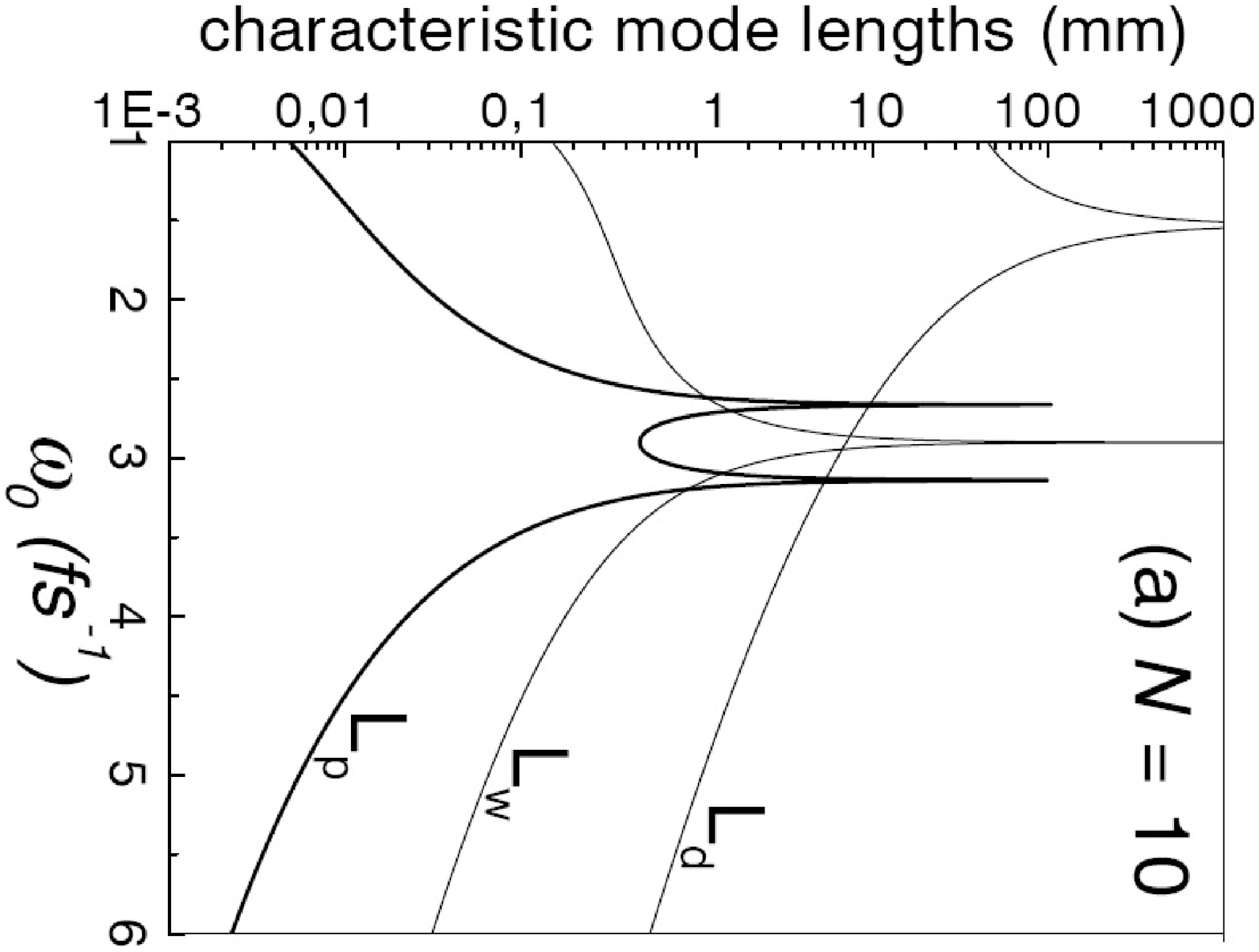}
\includegraphics[angle=90,width=4.2cm]{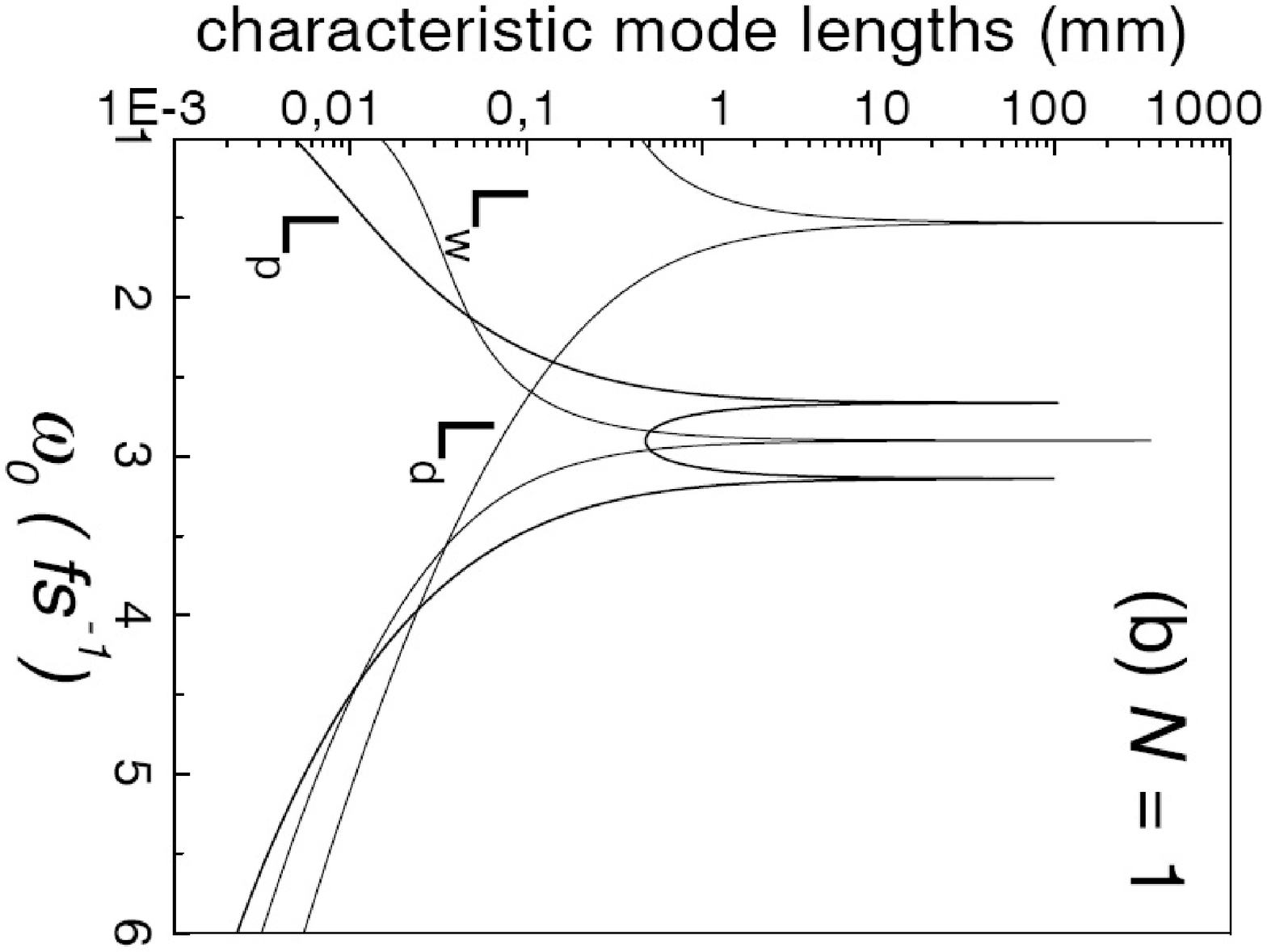}
\end{center}
\caption{\label{newfig2} Characteristic lengths of extraordinary second-harmonic wave modes of
different frequencies in the visible range that travel at the phase and group velocities of the
ordinary fundamental waves in LBO at room temperature. Mode bandwidths are
$\Delta\Omega/\omega_0=1/2\pi N$, with (a) $N=10$ and (b) $N=1$.}
\end{figure}

For illustration, we have evaluated the characteristic lengths of wave modes of different
frequencies $\omega_0$ that propagate in LBO at the phase and group velocities of the corresponding
fundamental waves of half-frequency. In Figs. \ref{newfig2}, the bandwidths $\Delta\Omega =
\omega_0/2\pi N$ correspond to ``$N$-cycle" pulses [duration $\sim(\Delta\Omega)^{-1}=N T_0$,
$T_0=2\pi/\omega_0$ period] at each frequency $\omega_0$. The value $N=10$ in Fig. \ref{newfig2}(a)
leads to a pulse duration $(\Delta\Omega)^{-1}\sim$ 20 fs at $\omega_0= 3.55$ fs$^{-1}$
($\lambda=0.53 \,\mu$m), of the same order as in previous experiments and numerical simulations.
Fig. \ref{newfig2}(b) shows, in contrast, the extreme case of ``single-cycle" wave modes. Generally
speaking, modes of long enough duration belong to, or participate mostly of, the PM-dominated case
[as in Fig.\ref{newfig2}(a) for most frequencies], modes of some (still unspecified) intermediate
duration belong to the GVM-dominated case, and extremely short modes to the GVD-dominated case,
since $L_p$ is independent on bandwidth, but $L_p$ and $L_d$ are inversely proportional to
$\Delta\Omega$ and $\Delta\Omega^{2}$, respectively. Depending, however, on the relative values of
$\alpha$, $\beta$ and $k_0^{\prime\prime}$ (particularly when one or two of them are very small),
the GVM-dominated case, even the PM-dominated case, can extend down to the single-cycle regime [as
in Fig.\ref{newfig2}(b) for most frequencies], or, on the contrary, the GVM-dominated case, even the
GVD-dominated case, apply to considerably long modes [as in the vicinity of the two singularities of
the $L_p$-curve of Fig. \ref{newfig2}(a)].

\subsection{Phase-mismatch-dominated case: Pulsed Bessel beam type modes}

Consider first modes with $|L_p|\ll |L_w|, |L_d|$. When $L_p>0$, the dispersion curve within the
spectral bandwidth can be approached by the real constant value $K(\Omega)\simeq
(2k_0L_p^{-1})^{1/2}$, or,
\begin{equation}\label{DISPB}
K(\Omega)\simeq \sqrt{2k_0\beta}\;\; (\mbox{if $\beta>0$}) ,
\end{equation}
[see Fig. \ref{fig3}(a)] regardless the exact dispersion curve is an actual hyperbola or ellipse [as
in Fig. \ref{fig3}(b)], that is, independently of the sign of material group velocity dispersion.
Wave modes under these conditions can only have superluminal phase velocity ($\beta>0$), but super-
or subluminal group velocity ($\alpha>0$ or $\alpha<0$, respectively), and will adopt, from Eqs.
(\ref{MODESI}) and (\ref{DISPB}), the approximate factorized form
\begin{equation}\label{PBB}
\Phi_{\alpha,\beta}(r,\tau+\alpha z)\simeq f(\tau+\alpha z) J_0\left(\sqrt{2 k_0\beta}\,r\right)
\end{equation}
of a PBB of transversal size of the order of $(2k_0\beta)^{-1/2}$.

Figure \ref{fig3}(c) shows the prototype PBB of this kind of wave modes [Eq. (\ref{PBB})] with a
Gaussian spectrum $\hat f(\Omega)$, that is, the limiting case $|L_p/L_w| = 0$, $|L_p/L_d|= 0$, or
horizontal thick lines of Figs. \ref{fig3}(a) and (b). In Fig. \ref{fig3}(d) we show, for
comparison, the wave mode with $|L_p/L_w| = 0.25$, $|L_p/L_d|=0.25$ and with the same Gaussian
spectrum, obtained numerically from Eq. (\ref{MODESI}). We see that the wave mode preserves a
spatiotemporal structure similar to that of the prototype PBB of Fig. \ref{fig3}(c), even if $|L_p|$
is not much smaller, but simply smaller than $|L_w|$ and $|L_d|$. Small differences can be
understood as incipient focus wave mode and O-wave type behavior, as described in the following
sections.

\subsection{Group-velocity-mismatch-dominated case: Envelope focus wave modes}

\begin{widetext}

\begin{figure}
\begin{center}
\begin{minipage}{5cm}
\includegraphics[angle=90,width=5cm]{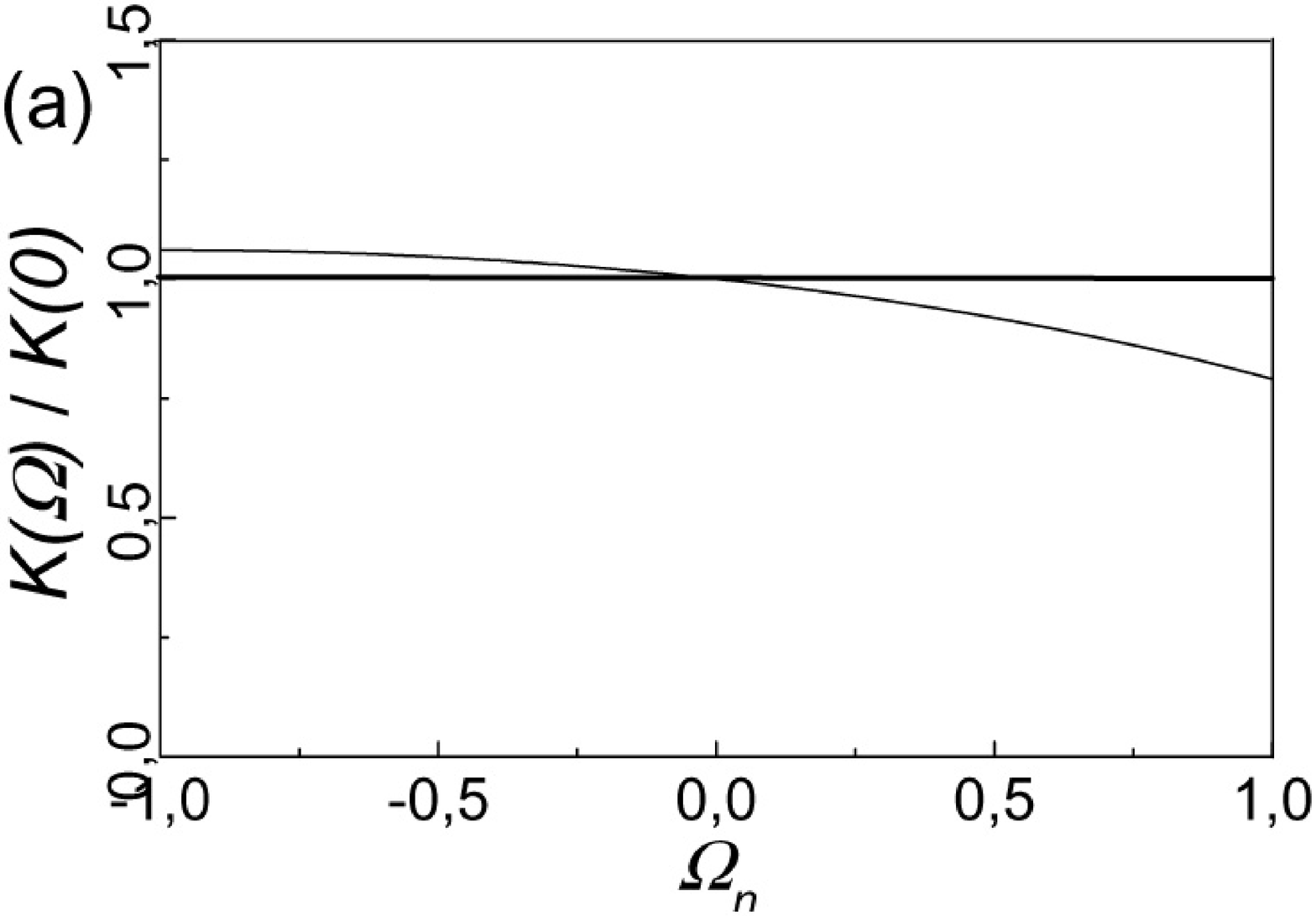}\\
\includegraphics[angle=90,width=5cm]{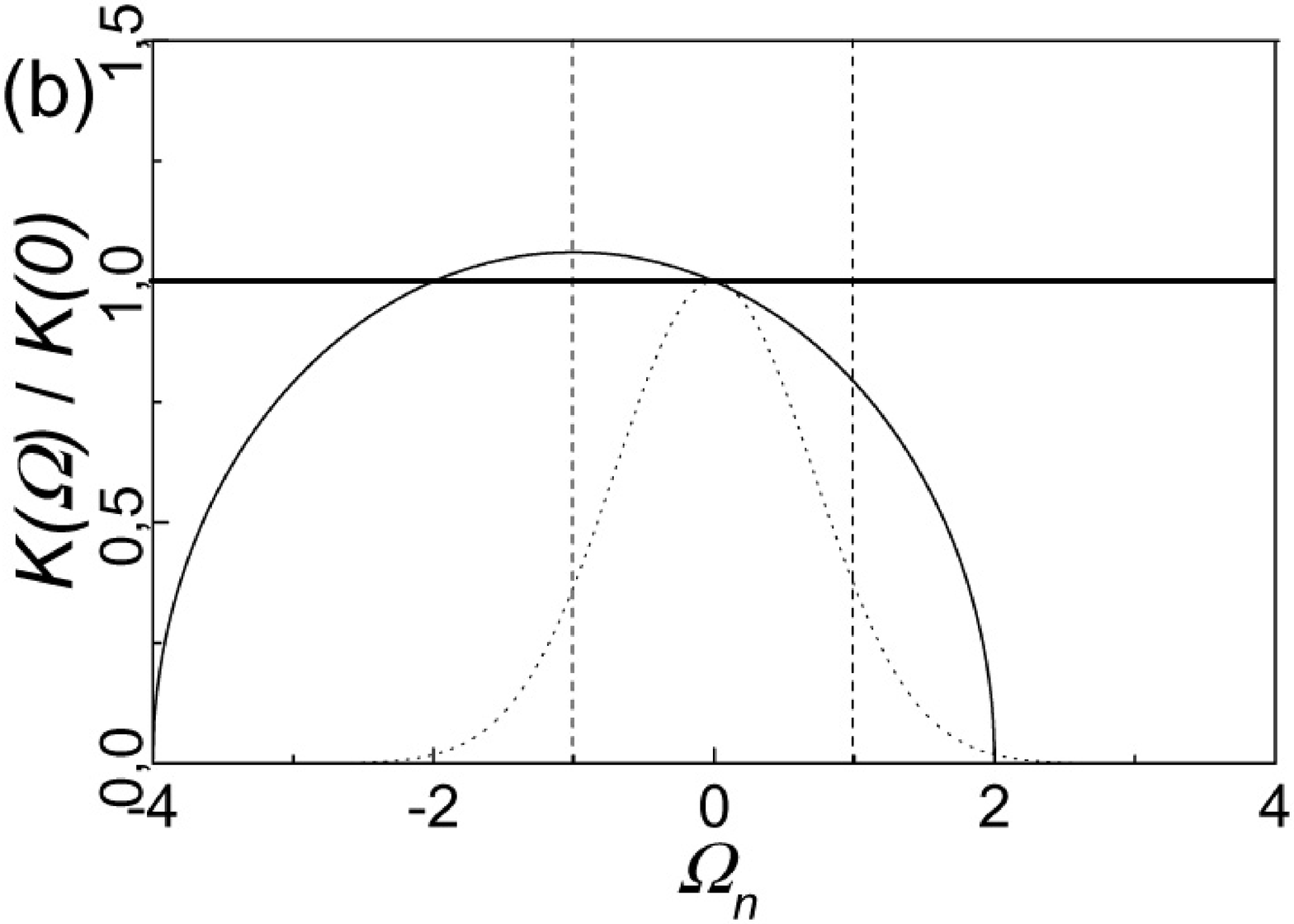}
\end{minipage}
\begin{minipage}{11cm}
\includegraphics[width=11cm]{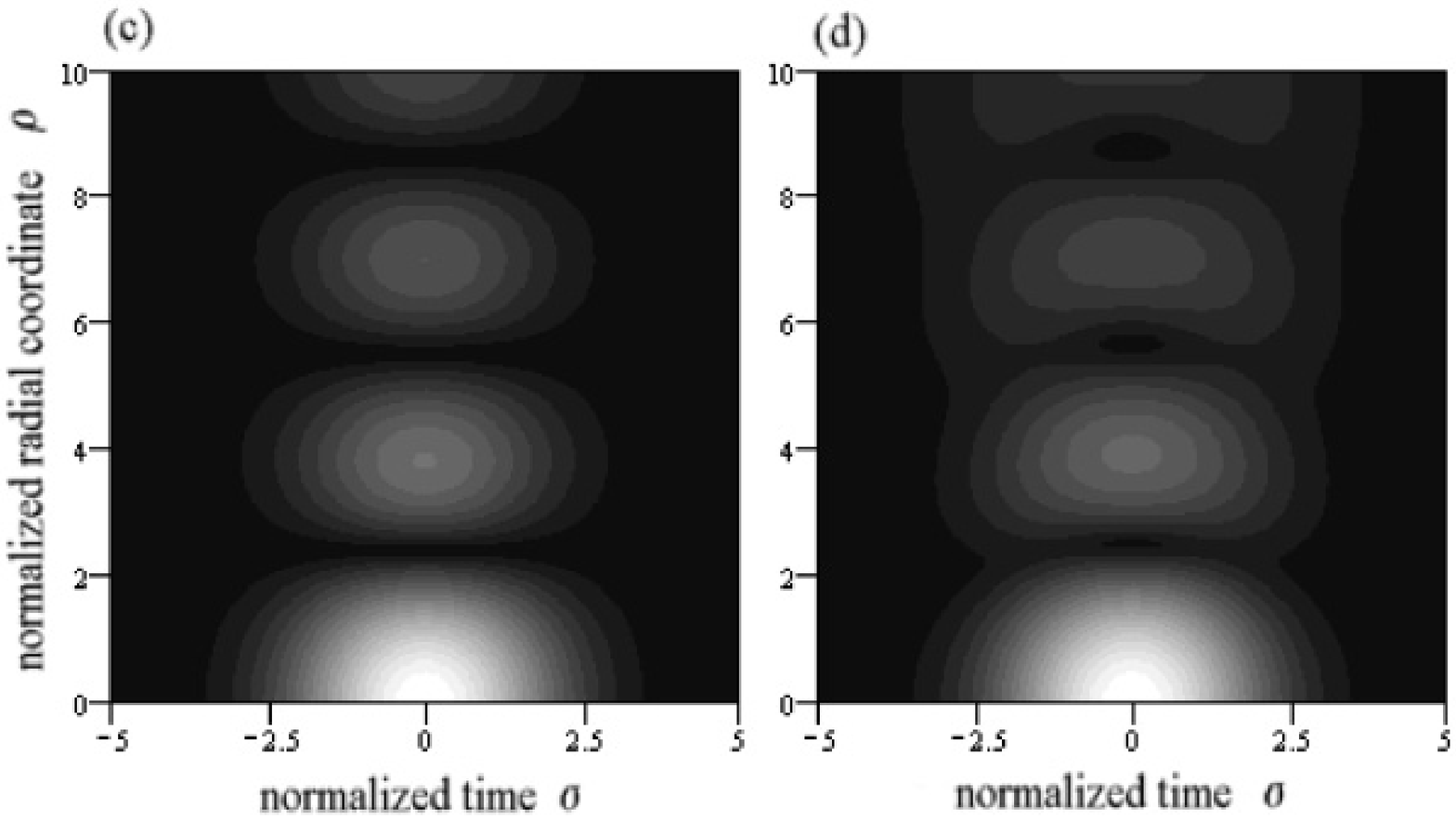}
\end{minipage}
\end{center}
\caption{\label{fig3} (a) Dispersion curve within the bandwidth for $|L_p|/|L_w|\rightarrow 0$,
$|L_p|/|L_d|\rightarrow 0$ (thick curve), and for $L_p/L_w=-0.25$, $L_p/L_d=-0.25$ (thin curve). (b)
The same as in (a) but also outside the bandwidth of the Gaussian spectrum (in arbitrary units)
$\hat f(\Omega)= \exp[-(\Omega/\Delta\Omega)^2]$. (c) and (d) Gray-scale plots of the amplitude
$|\Phi_{\alpha,\beta}|$ of (c) the PBB of Eq. (\ref{PBB}) with spectrum $\hat
f(\Omega)=\exp[-(\Omega/\Delta\Omega)^2]$ (i.e., $f(\tau)\propto\exp[-(2\Delta\Omega \tau)^2]$) and
(d) of the mode with $L_p/L_w=-0.25$, $L_p/L_d=-0.25$ and same spectrum as in (c), numerically
calculated from Eq. (\ref{MODESI}). Normalized coordinates are $\sigma=(\tau+\alpha z)\Delta\Omega$,
$\rho=r/r_0$, with $r_0=(2k_0\beta)^{-1/2}$.}
\end{figure}

\begin{figure}[t!]
\begin{center}
\begin{minipage}{5cm}
\includegraphics[width=5cm]{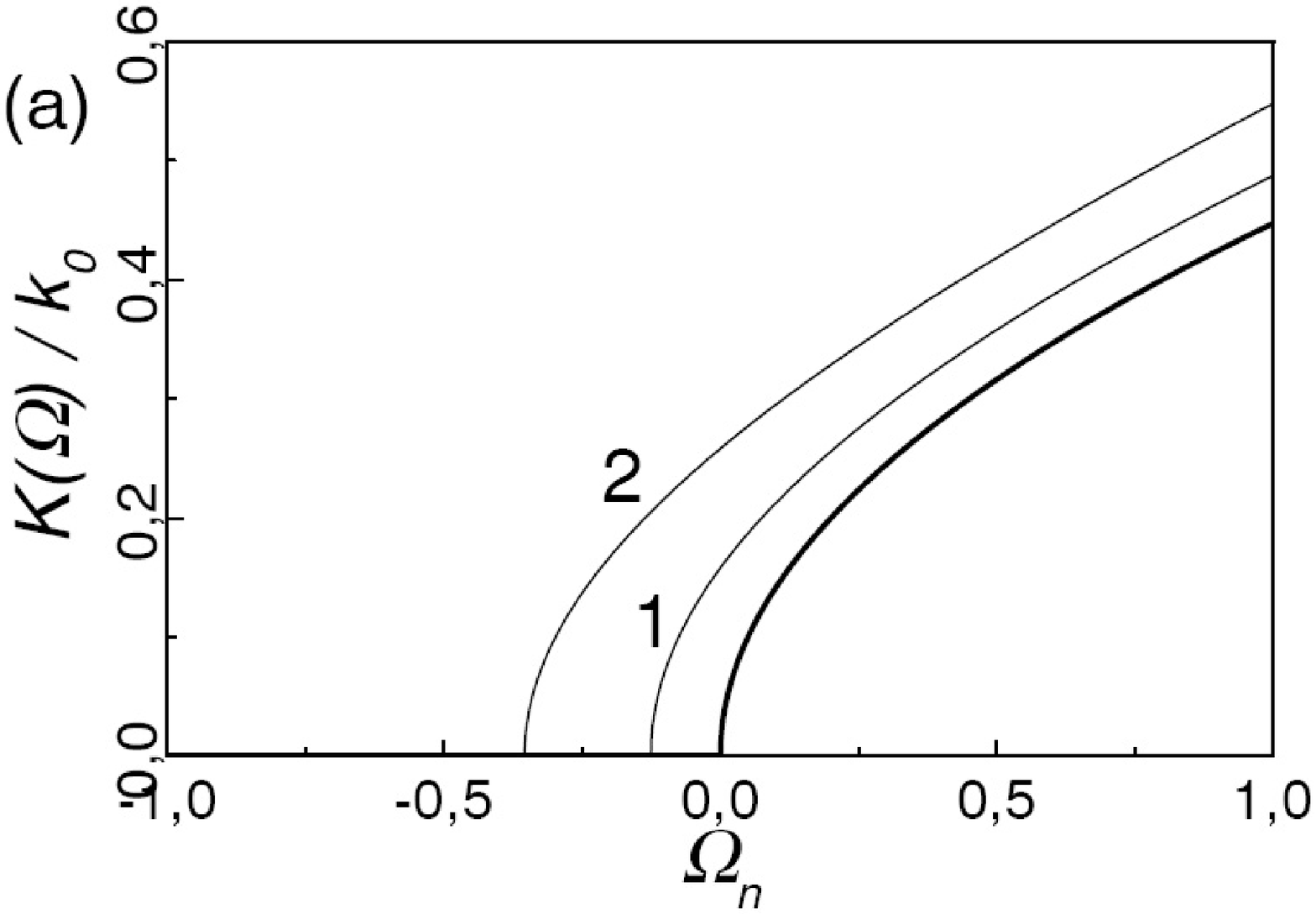}\\
\includegraphics[angle=90,width=5cm]{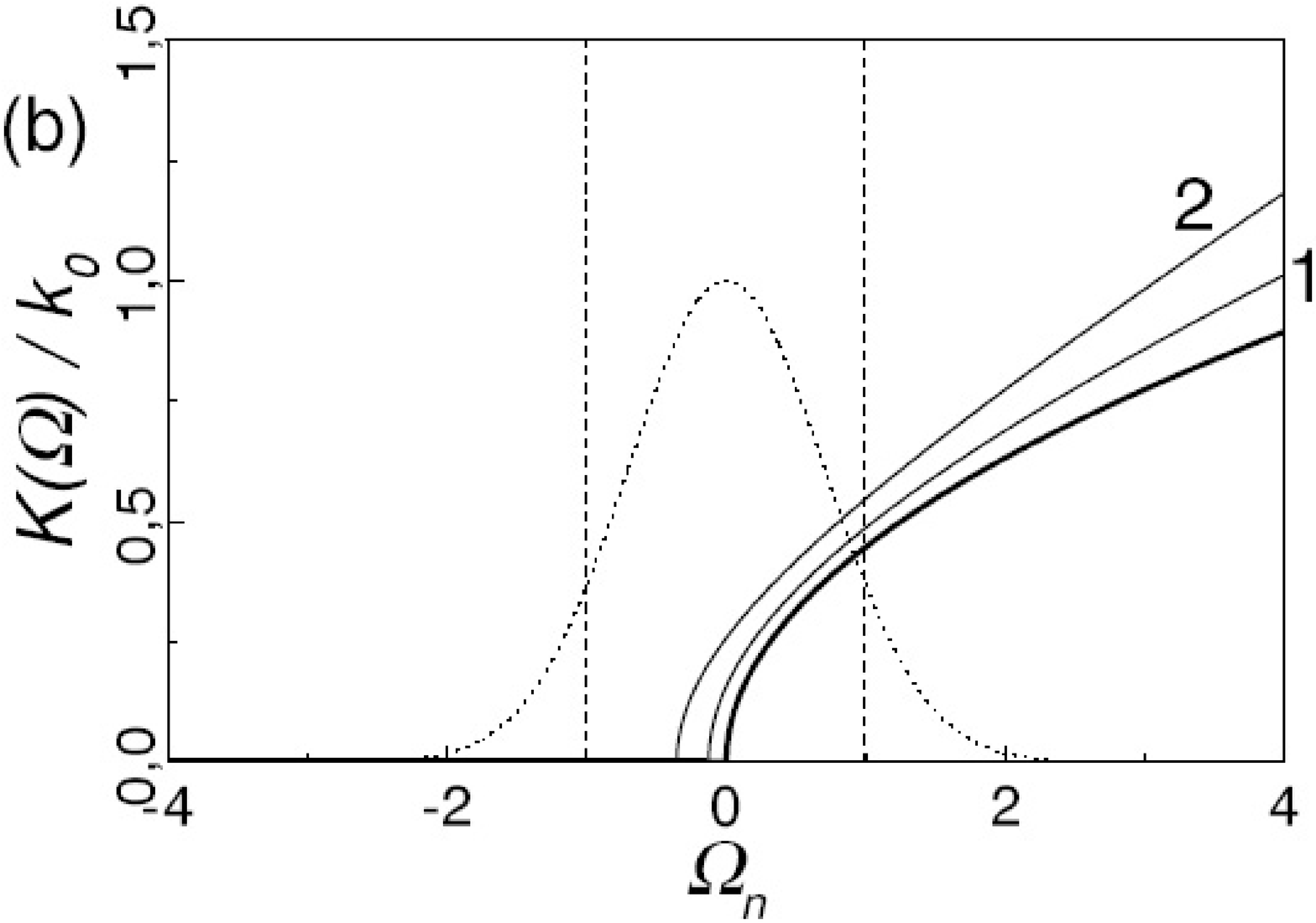}
\end{minipage}
\begin{minipage}{11cm}
\includegraphics[width=11cm]{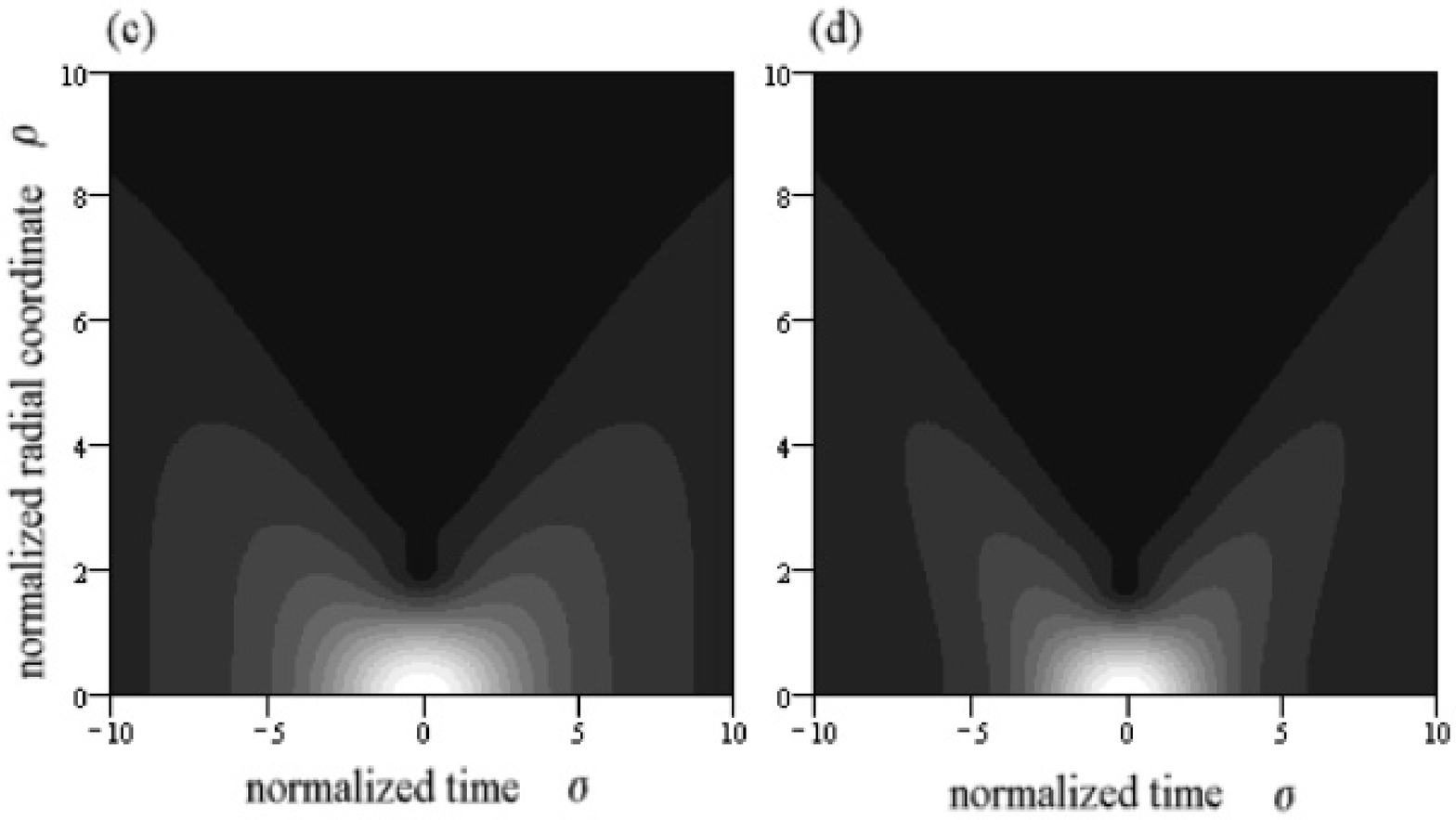}
\end{minipage}
\end{center}
\caption{\label{fig4} (a) Dispersion curve within the bandwidth for $L_w=10/k_0$,
$L_w/L_p\rightarrow 0$, $L_w/L_d\rightarrow 0$ (thick curve), for $L_w=10/k_0$, $L_w/L_p=1/8$,
$L_w/L_d=1/8$ (thin curve, label 1), and for $L_w=10k_0$, $L_w/L_p=1/3$, $L_w/L_d=1/3$ (thin curve,
label 2). (b) The same as in (a) but also outside the bandwidth of the Gaussian spectrum (in
arbitrary units) $\hat f(\Omega)= \exp[-(\Omega/\Delta\Omega)^2]$. (c) and (d) Gray-scale plots of
the amplitude $|\Phi_{\alpha,\beta}|$ of (c) the prototype eFWM [thick dispersion curve in (a)] with
spectrum $\hat f(\Omega)=\exp[-(\Omega/\Delta\Omega)^2]$, and of (d) of the mode with $L_w/L_p=1/3$,
$L_w/L_d=1/3$ [thin dispersion curve 2 in (a)], numerically calculated from Eq. (\ref{MODESI}).
Normalized coordinates are $\sigma=(\tau+\alpha z)\Delta\Omega$, $\rho=r/r_0$, with
$r_0=(2/k_0\Delta\Omega|\alpha|)^{1/2}$.}
\end{figure}

\end{widetext}

The case $|L_w|\ll |L_p|, |L_d|$ leads to a new kind of wave modes that has not been reported. The
dispersion curve within the bandwidth is now of the form of the horizontal parabola $K(\Omega)\simeq
(2k_0L_w^{-1}\Omega_n)^{1/2}$ with vertex at $\Omega=0$, or,
\begin{equation}
K(\Omega)\simeq\sqrt{2k_0\alpha\Omega} , \label{DISPFWM}
\end{equation}
[see Fig. \ref{fig4}(a)], regardless material dispersion is normal [as in Fig. \ref{fig4}(b)] or
anomalous. For modes with superluminal group velocity ($\alpha>0$), the horizontal parabola is
right-handed [as in Figs. \ref{fig4}(a) and (b)], and left-handed for subluminal modes ($\alpha<0$).
Independently of the group velocity, phase velocity can be superluminal ($\beta>0$) or subluminal
($\beta<0$). In any case, their spatiotemporal form can be approached by Eq. (\ref{MODESI}) with
$K(\Omega)$ given by Eq. (\ref{DISPFWM}). Moreover, with the two-sided exponential spectrum $\hat
f(\Omega)=(2\pi/\Delta\Omega)\exp(-|\Omega|/\Delta\Omega)$, Eq. (\ref{MODESI}) yields
\begin{equation}\label{PEFWM}
\Phi_{\alpha,\beta}(r,\tau+\alpha z) \simeq \frac{-i\tau_0}{\tau+\alpha z - i\tau_0}
\exp\left[\frac{ik_0|\alpha| r^2}{2(\tau+\alpha z - i\tau_0)}\right]
\end{equation}
for superluminal modes ($\alpha>0$), and the complex conjugate of the r.h.s. of Eq.~(\ref{PEFWM})
for subluminal modes ($\alpha<0)$. In Eq. (\ref{PEFWM}), $\tau_0\equiv(\Delta\Omega)^{-1}$
characterizes the mode duration. The mode spot size at pulse center ($\tau+\alpha z=0$) can be
characterized by $r_0=(2/k_0\Delta\Omega|\alpha|)^{1/2}$.

The functional form of the reduced envelope in Eq. (\ref{PEFWM}) is similar to the fundamental
Brittigham-Ziolkowski focus wave mode (FWM) \cite{BRI,ZIOL}, and as such will be called envelope
focus wave mode (eFWM). There are, however, important physical differences between them, which can
be understood for the respective expressions of the complete fields $E$ of both kind of waves,
namely,
\begin{eqnarray}
E_{\alpha,\beta}(r,z,t)& \simeq & \frac{-i\tau_0}{\tau+\alpha z - i\tau_0}
\exp\left[\frac{ik_0|\alpha| r^2}{2(\tau+\alpha z - i\tau_0)}\right] \nonumber \\
                       & \times & \exp(-i\beta z)\exp(-i\omega_0 t +i k_0z) ,
\end{eqnarray}
for the envelope focus wave mode,
\begin{equation}
E(r,z,t)=\frac{-i\tau_0}{\tau-i\tau_0}\exp\left[\frac{ik_0r^2}{2c(\tau-i\tau_0)}\right]
\exp(-i\omega_0t-ik_0z) ,
\end{equation}
with $k_0=\omega_0/c$, for the fundamental FWM \cite{ZIOL}. The fundamental FWM is a localized,
stationary {\em free-space} wave whose envelope propagates at luminal group velocity $c$, whereas
the carrier oscillations back-propagate at the same velocity $c$. The eFWM is also a stationary,
localized wave with the same intensity distribution as the fundamental FWM, but propagates in a {\em
dispersive medium} with super- or subluminal group velocity $1/(k'_0-\alpha)$. The carrier
oscillations propagate in the same direction at super- or subluminal phase velocity
$\omega_0/(k_0-\beta)$.

Figure \ref{fig4}(c) shows the prototype eFWM of this kind of wave modes, obtained from numerical
integration of Eq. (\ref{MODESI}) with the approximate dispersion curve
$K(\Omega)=\sqrt{2k_0\alpha\Omega}$ [thick curves in Figs. \ref{fig4}(a) and (b)] i.e., in the
limiting case $|L_w/L_p| = 0$, $|L_w/L_d|= 0$), and a Gaussian spectrum. To pursue the validity of
the model eFWM to describe this kind of wave modes, we have also evaluated the wave mode field in
some non-limiting cases with the same Gaussian spectrum. For $|L_w/L_p| = 1/8$, $|L_w/L_d|=1/8$
[thin curves in Figs. \ref{fig4}(a) and (b), label 1], the mode is nearly undistinguishable from the
prototype eFWM, despite the dispersion curve differs significantly from the limiting one. Even for
the relatively large ratios $|L_w/L_p|=1/3$, $|L_w/L_d|=1/3$ [thin curves in Figs. \ref{fig4}(a) and
(b), label 2], the calculated wave mode [see Fig. \ref{fig4}(d)] exhibits the same eFWM structure,
with some incipient eX-wave behavior because of the actual hyperbolic form (not parabolic) of the
dispersion curve, as explained in the next section.

\subsection{Group-velocity-dispersion-dominated case: Envelope X and envelope O type modes}

\subsubsection{Normal group velocity dispersion: Envelope X waves}

\begin{widetext}

\begin{figure}[t!]
\begin{center}
\includegraphics[width=5cm]{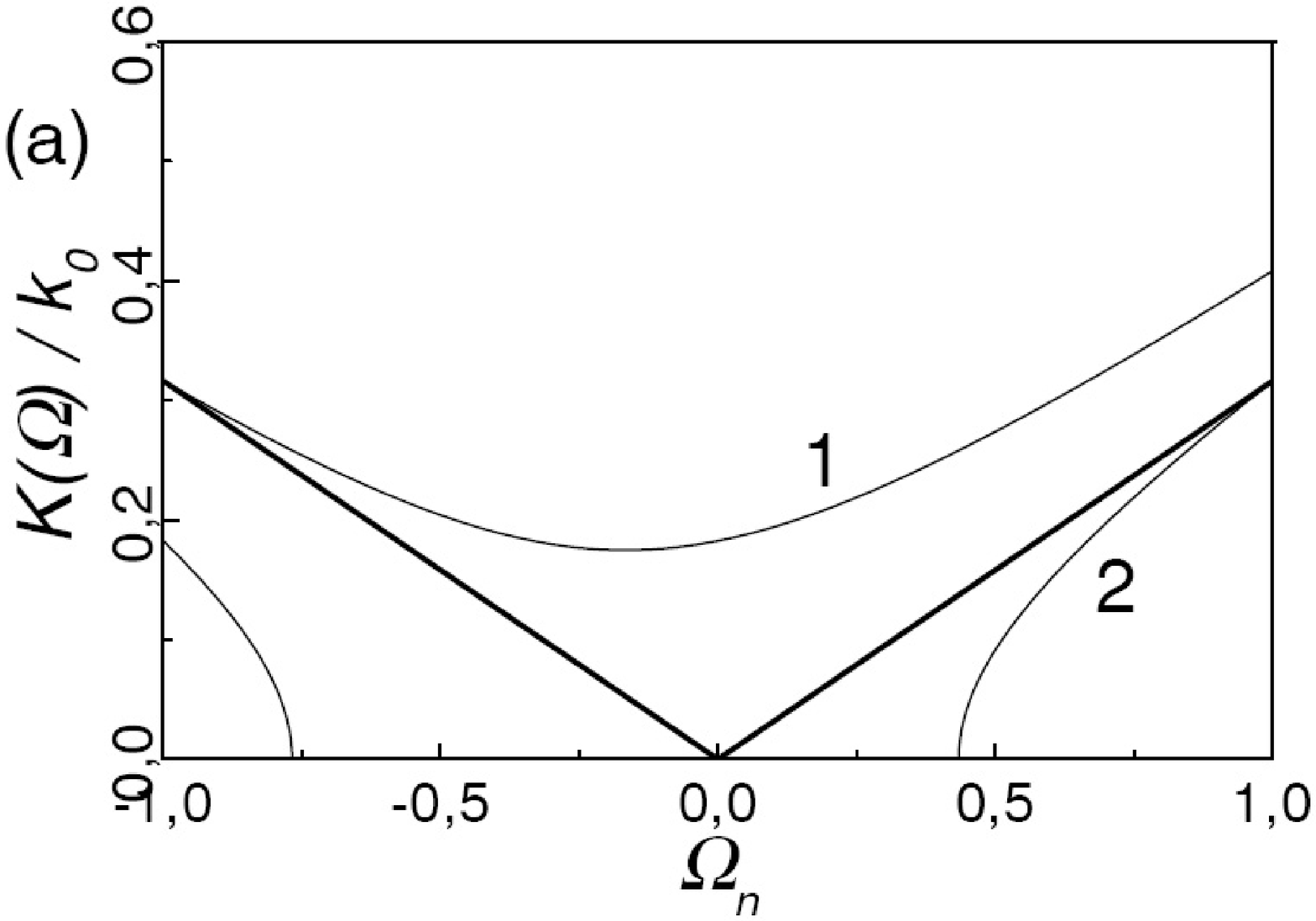}
\includegraphics[angle=90,width=5cm]{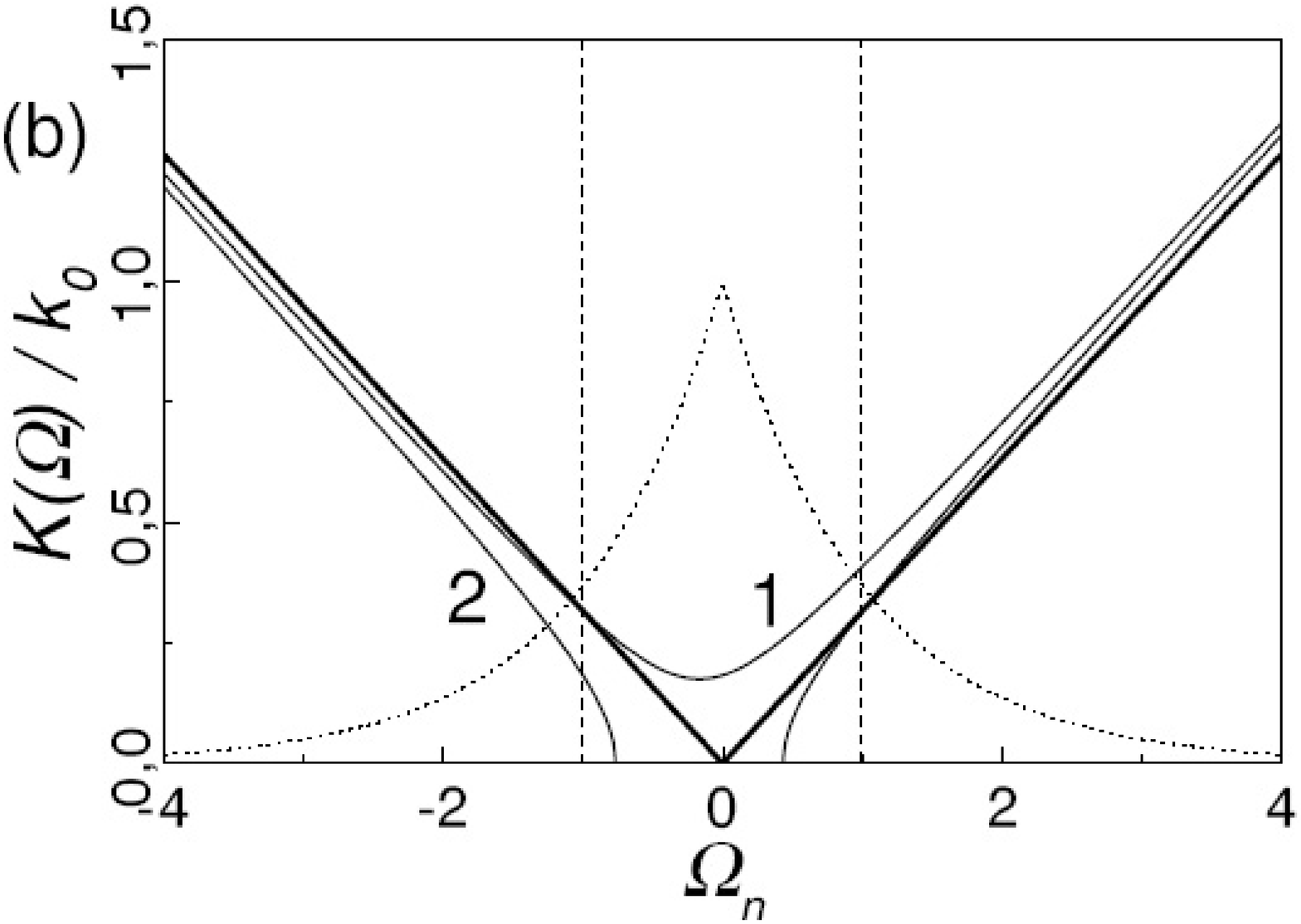}
\end{center}
\begin{center}
\includegraphics[width=16cm]{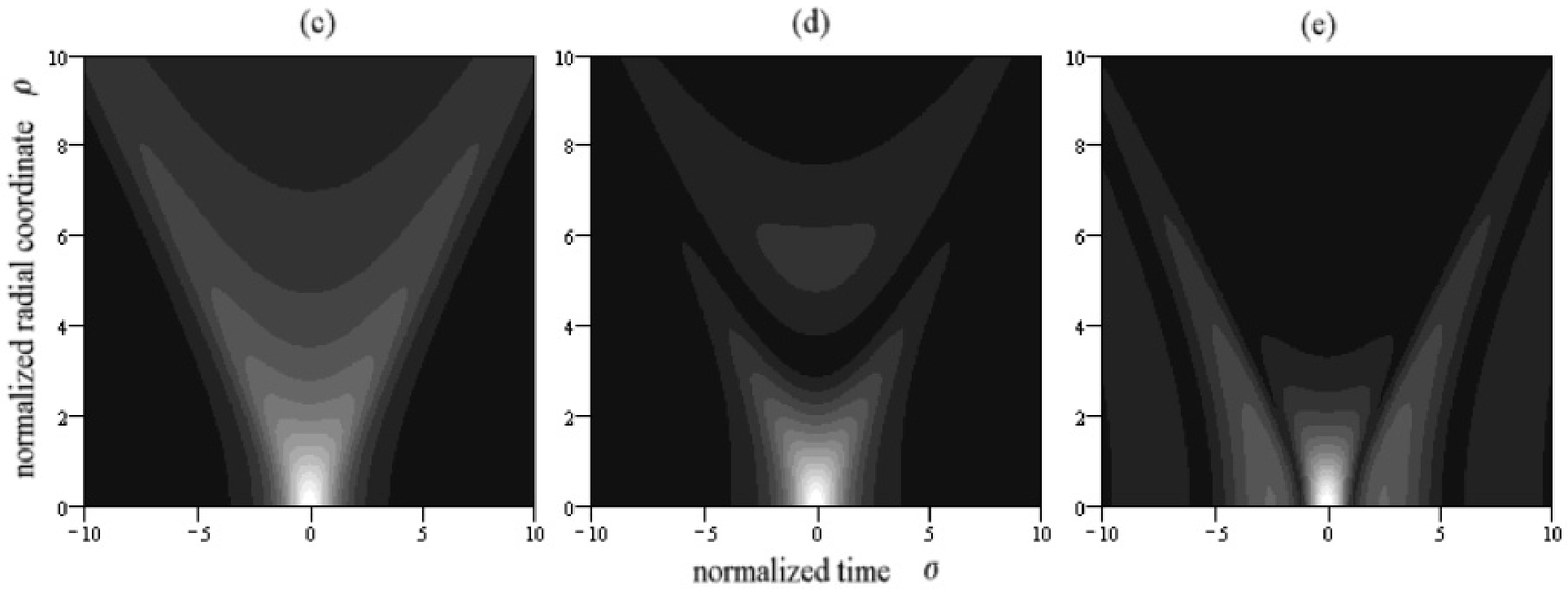}
\end{center}
\caption{\label{fig5} (a) Dispersion curve within the bandwidth for $L_d=10/k_0$,
$L_d/L_p\rightarrow 0$, $L_d/L_w\rightarrow 0$, with $L_d>0$ (thick curve), for $L_d=10/k_0$,
$L_d/L_p=1/6$, $L_d/L_w=1/6$ (thin curve, label 1), and for $L_d=10k_0$, $L_d/L_p=-1/6$,
$L_d/L_w=1/6$ (thin line, label 2). (b) The same as in (a) but also outside the bandwidth of the
spectrum (in arbitrary units) $\hat f(\Omega)= \exp(-|\Omega|/\Delta\Omega)$. (c-e)  Gray-scale
plots of the amplitude $|\Phi_{\alpha,\beta}|$ of (c) the prototype eX [thick dispersion curve in
(a)] with exponential spectrum $\hat f(\Omega)=\exp[-(\Omega/\Delta\Omega)^2]$, (d) of the mode with
$L_d/L_p=1/6$, $L_d/L_w=1/6$ [thin dispersion curve 1 in (a)], and (e) of the mode with
$L_d/L_p=-1/6$, $L_d/L_w=1/6$ [thin dispersion curve 2 in (a)]. Normalized coordinates are
$\sigma=(\tau+\alpha z)\Delta\Omega$, $\rho=r/r_0$, with
$r_0=(k_0k_0^{\prime\prime}\Delta\Omega^2)^{-1/2}$.}
\end{figure}

\end{widetext}

We consider finally modes with $|L_d|\ll |L_p|,|L_w|$, or modes of short enough duration, or
propagating in a medium with large enough GVD. When material dispersion is normal
($k_0^{\prime\prime}>0$), the dispersion curve within the bandwidth approaches the X-shaped curve
[see Fig. \ref{fig5}(a)]
\begin{equation}\label{DISPX}
K(\Omega)\simeq \sqrt{k_0k_0^{\prime\prime}}|\Omega|
\end{equation}
of the limiting case $|L_d/L_p|, |L_d/L_w| = 0$. The actual dispersion curve of a mode may be
slightly shifted towards negative frequencies [as in Figs. \ref{fig5}(a) and (b), labels 1 and 2] or
positive frequencies for modes with superluminal ($\alpha>0$) or subluminal ($\alpha<0$) group
velocity, respectively. For modes with superluminal phase velocity ($\beta>0$), $K(\Omega)$ is real
everywhere [Fig. \ref{fig5}(b), label 1], but for modes with subluminal phase velocity there is a
narrow frequency gap about $\Omega=0$ [Fig. \ref{fig5}(b), label 2]. A prototype wave mode for this
case can be obtained by introducing the approximate dispersion curve of Eq. (\ref{DISPX}) into Eq.
(\ref{MODESI}). With the two-side exponential spectrum $\hat
f(\Omega)=(2\pi/\Delta\Omega)\exp(-\Omega/\Delta\Omega)$ we obtain
\begin{equation}\label{EX}
\Phi_{\alpha,\beta}(r,\tau+\alpha z)\simeq \Re \left\{\frac{\tau_0}{\sqrt{k_0k_0^{\prime\prime} r^2
+ [\tau_0+i(\tau+\alpha z)^2]}}\right\} ,
\end{equation}
where $\tau_0\equiv(\Delta\Omega)^{-1}$ measures the pulse duration. Equation (\ref{EX}) is the eX
wave recently described in Ref. \cite{POOL2003} as an exact, stationary and localized solution of
the paraxial wave equation with luminal phase and group velocities ($\alpha=\beta=0$) in media with
normal GVD. The eX wave (\ref{EX}) is understood here as an approximate expression for modes with
$\alpha, \beta$ such that $|L_d/L_p|\ll 1$, $|L_d/L_w|\ll 1$. The spatiotemporal form of the eX wave
is shown in Fig. \ref{fig5}(c). For $L_d/L_p = 1/6$ ($\beta>0$), $L_d/L_w=1/6$ [thin curves in Figs.
\ref{fig5}(a) and (b), label 1], the mode retains an X-shaped structure [Fig. \ref{fig5}(d)] despite
the dispersion curve differ significantly from the limiting one. Incipient PBB behavior, or radial
oscillations, originates from the nearly horizontal dispersion curve in the central part of the
spectrum. For $L_d/L_p = -1/6$ ($\beta<0$), $L_d/L_w=1/6$ [thin curves in Figs. \ref{fig5}(a) and
(b), label 2], the X-shaped mode [Fig. \ref{fig5}(d)] shows instead incipient eFWM behavior (light
is within the cone), together with temporal oscillations arising from the frequency gap in the
dispersion curve.

\subsubsection{Anomalous group velocity dispersion: Envelope O waves}

\begin{widetext}

\begin{figure}
\begin{center}
\begin{minipage}{11cm}
\includegraphics[width=11cm]{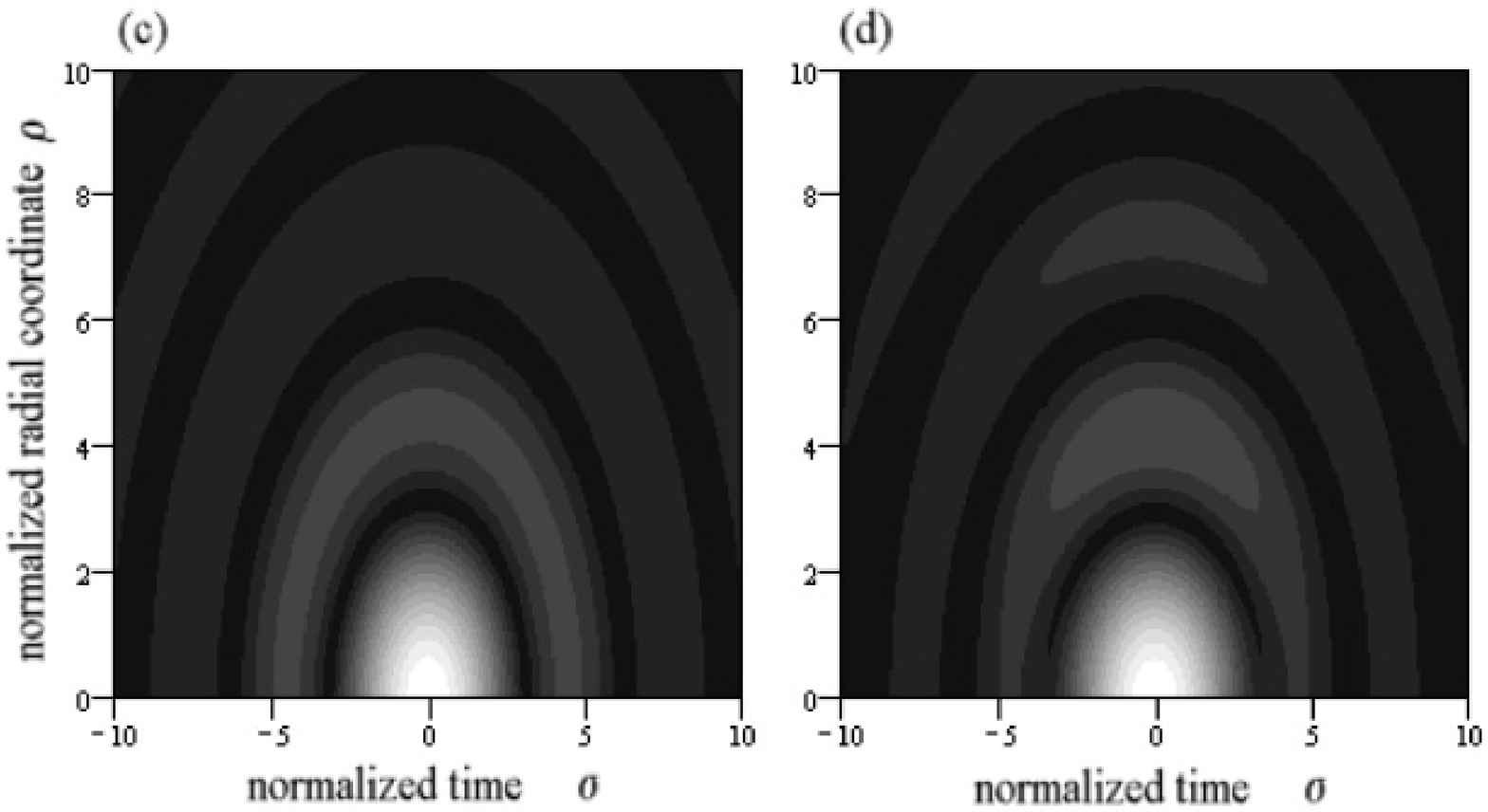}
\end{minipage}
\begin{minipage}{5cm}
\includegraphics[width=5cm]{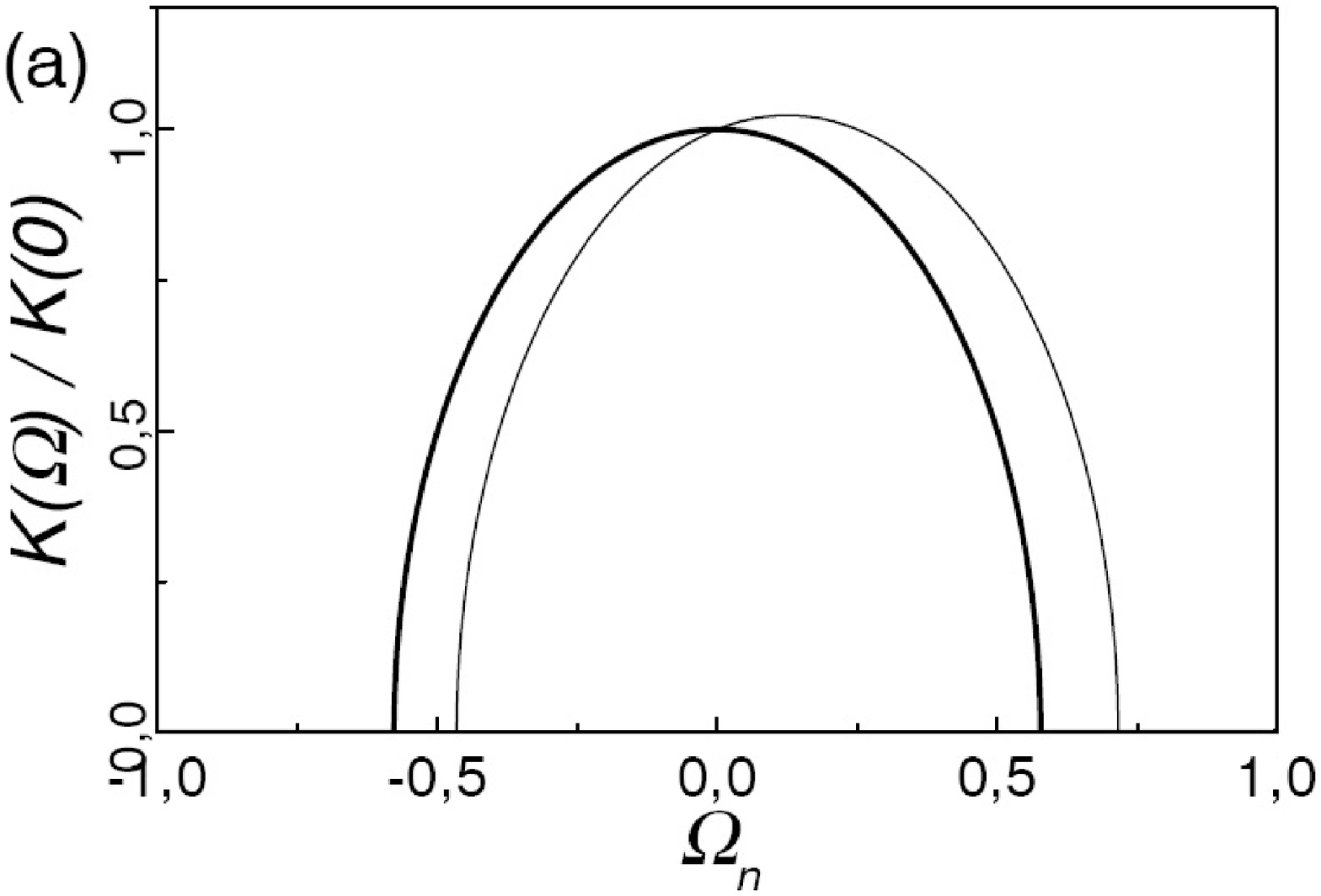}\\
\includegraphics[angle=90,width=5cm]{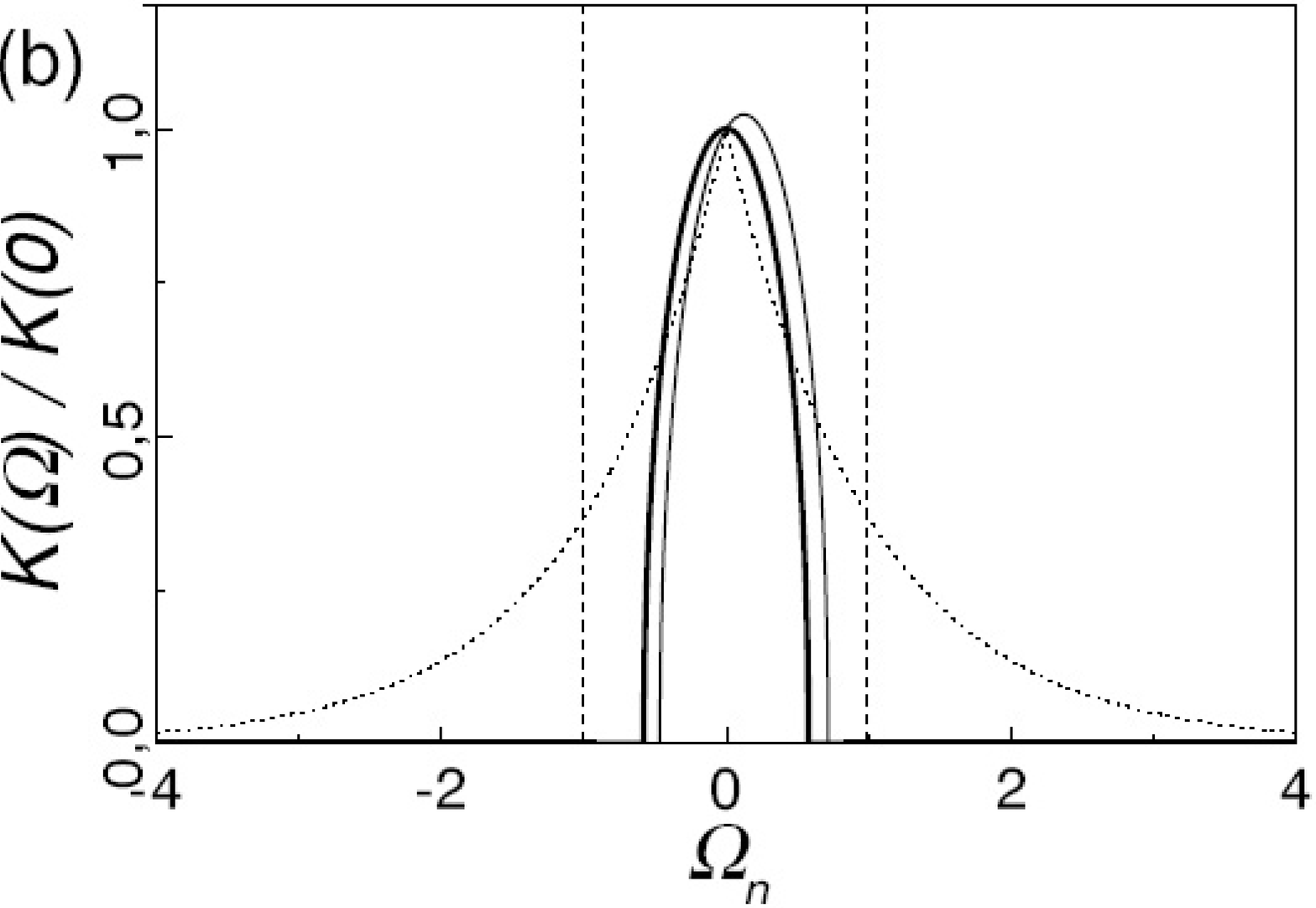}
\end{minipage}
\end{center}
\caption{\label{fig6} (a) Dispersion curve within the bandwidth for $|L_d|/|L_p|\rightarrow 0$,
$|L_d|/|L_w|\rightarrow 0$, with $L_d<0$ (thick line), and for $L_d/L_p=-1/6$, $L_d/L_w=-1/8$ (thin
curve). (b) The same as in (a) but also outside the bandwidth of the spectrum (in arbitrary units)
$\hat f(\Omega)= \exp(-|\Omega|/\Delta\Omega)$. (c) Gray-scale plot of the amplitude
$|\Phi_{\alpha,\beta}|$ of the eO wave of Eq. (\ref{EOW}), (d) and of the mode with $L_d/L_p=-1/6$,
$L_d/L_w=-1/8$ and the exponential spectrum of (b), numerically calculated from Eq. (\ref{MODESI}).
Normalized coordinates are $\sigma=(\tau+\alpha z)\sqrt{2\beta/|k_0^{\prime\prime}|}$,
$\rho=\sqrt{2k_0\beta} r$.}
\end{figure}

\end{widetext}

When $|L_d|\ll |L_p|,|L_w|$ but GVD is anomalous, the dispersion curve within the bandwidth can be
approached by the ellipse centered on $\Omega=0$ [Figs. \ref{fig6}(a) and (b), thick curves] given
by the expression
\begin{equation}\label{DISPO}
K(\Omega)\simeq\sqrt{2k_0(\beta-|k_0^{\prime\prime}|\Omega^2/2)} .
\end{equation}
Note that the term with $\beta$, no matter how small it is, must be retained to reproduce the
real-valued part of the dispersion curve. The group velocity of the mode can be slightly subluminal
($\alpha<0$) or superluminal ($\alpha>0$), as in Fig. \ref{fig6}(a) and (b) (thin curves), but the
phase velocity of these modes is always superluminal ($\beta>0$). An approximate analytical
expression for this type of modes can be obtained by introducing the approximate dispersion curve of
Eq. (\ref{DISPO}) into Eq. (\ref{MODESI}). Under condition $|L_d|\ll |L_p|$, the frequency gap
$\Omega_g\simeq\sqrt{2\beta/|k_0^{\prime\prime}|}$ is much smaller than $\Delta\Omega$, so that the
amplitude spectrum $\hat f(\Omega)$ can be assumed to take a constant value in the integration
domain of integral in Eq. (\ref{MODESI}), which then yields the expression
\begin{eqnarray}\label{EOW}
\Phi_{\alpha,\beta} & \simeq & \frac{1}{\sqrt{k_0|k_0^{\prime\prime}| r^2 +(\tau\!+\!\alpha z)}}
\nonumber \\
                 &\times &  \sin \left[\sqrt{2\beta/|k_0^{\prime\prime}|}\sqrt{k_0|k_0^{\prime\prime}|
                 r^2\! +\!(\tau\!+\!\alpha z)}\right],
\end{eqnarray}
of the same form as the O-type impulse response mode in media with anomalous dispersion. Figure
\ref{fig6}(c) shows its spatiotemporal form. For comparison, the wave mode with $L_d/L_p=-1/6$,
$L_d/L_w=-1/8$ [Fig. \ref{fig6}(a), thin curve] and the two-sided exponential spectrum [Fig.
\ref{fig6} (b)] was calculated from Eq. (\ref{MODESI}), and its O-shaped spatiotemporal form is
depicted in Fig. \ref{fig6}(d).

\section{Nonparaxial descriptions of wave modes} \label{NP}

The purpose of this section is to show that the preceding classification of wave modes in dispersive
media in terms of the characteristic lengths remains essentially unaltered when performed from the
more exact nonparaxial approach, if condition (\ref{COND1}) of quasi-monochromaticity, and
(\ref{COND2}) and (\ref{COND3}) of quasi-luminal group and phase velocities are satisfied.

We consider now the polychromatic Bessel beam,
\begin{eqnarray}
E(r,z,t)&=& \frac{1}{2\pi}\int_{\mbox{$K$ real}} d\omega \hat f(\omega-\omega_0) \nonumber \\
        &\times& J_0(K r)
\exp(ik_z z)\exp(-i\omega t), \label{POLI}
\end{eqnarray}
where $K$ and $k_z$ must be related by $K=\sqrt{k^2(\omega)-k_z^2}$ for each monochromatic Bessel
beam component to satisfy the Helmholtz equation $\Delta \hat E+ k^2(\omega)\hat E=0$. Stationarity
of the intensity in some moving reference frame requires the axial propagation constant $k_z$ to be
a linear function of frequency \cite{SO96}, a condition that is suitably expressed as
\begin{equation}
k_z(\Omega)=(k_0-\beta) + (k'_0-\alpha)\Omega .
\end{equation}
Equation (\ref{POLI}) can be then rewritten in the form $E(r,z,t)=\Phi_{\alpha,\beta}(r,\tau+\alpha
z )\exp(-i\beta z)\exp(-i\omega_0 t + ik_0 z)$, where the reduced envelope is given by the same
expression as in the paraxial case, namely,
\begin{eqnarray}\label{NPMODES}
\Phi_{\alpha,\beta}(r,\tau+\alpha z)&=&\frac{1}{2\pi}\int_{\mbox{$K(\Omega)$ real}} d\Omega \hat
f(\Omega) \nonumber \\
          &\times & J_0[K(\Omega)r]\exp[-i\Omega(\tau+\alpha z)] ,
\end{eqnarray}
but with a transversal dispersion relation $K(\Omega)=\sqrt{k^2(\Omega)-k_z^2(\Omega)}$ given now by
\begin{eqnarray}
K(\Omega)&=& \left[ (2k_0\beta -\beta^2) + 2(k_0\alpha+k'_0\beta-\alpha\beta)\Omega \right.\nonumber \\
         &+& \left.
(k_0k_0^{\prime\prime} + 2k'_0\alpha-\alpha^2)\Omega^2\right]^{1/2} \label{NPDISP}
\end{eqnarray}
up to second order in dispersion [$k(\Omega)=k_0+k_0'\Omega+ k_0^{\prime\prime}\Omega^2/2$].

In the case of propagation in free-space ($k_0=\omega_0/c$, $k'_0=1/c$, $k_0^{\prime\prime}=0$, with
$c$ the speed of light in vacuum), Eqs. (\ref{NPMODES}) and (\ref{NPDISP}) yield Eq. (7) of Ref.
\onlinecite{SAAJOSAA} for general free-space FWMs, if the identifications $\alpha=(1-\gamma)/c$ and
$\beta=\omega_0\alpha+2\gamma\beta_s$ are made ($\gamma$ and $\beta_s$ being the the free parameters
defined in Ref. \onlinecite{SAAJOSAA}). In particular, the case with $\alpha=0$ yields the original
Brittigham's FWM \cite{BRI,ZIOL}, and the case with $\beta = \omega_0\alpha$ yields the Bessel-X
pulse of cone angle $\theta=(2c\alpha)^{1/2}$, or X wave with narrow spectral amplitude centered at
an optical frequency, introduced by Saari in Ref. \onlinecite{SAALP}, and demonstrated in Ref.
\onlinecite{SAAPRL}.

\begin{figure}
\begin{center}
\includegraphics[width=4.2cm]{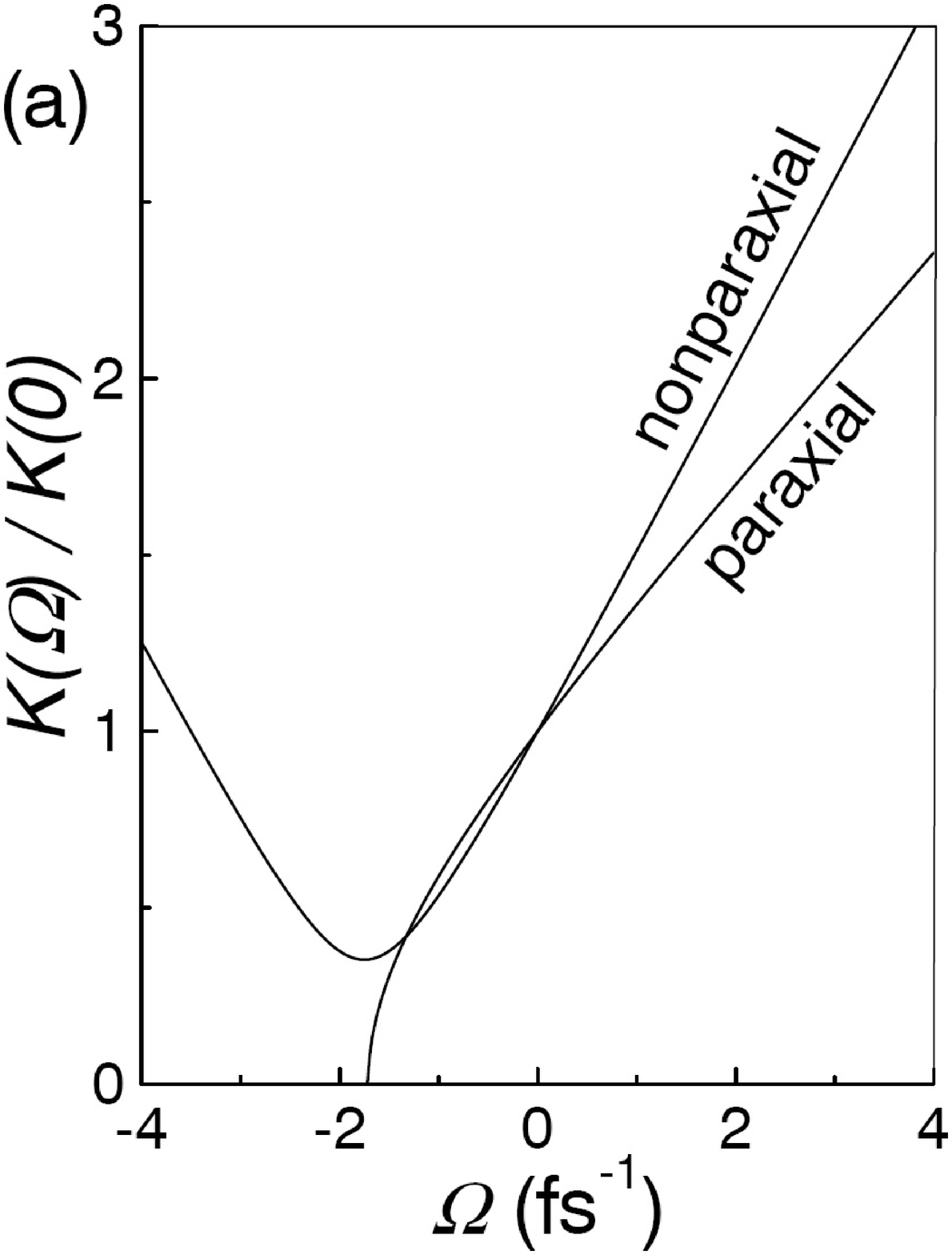}
\includegraphics[width=4.2cm]{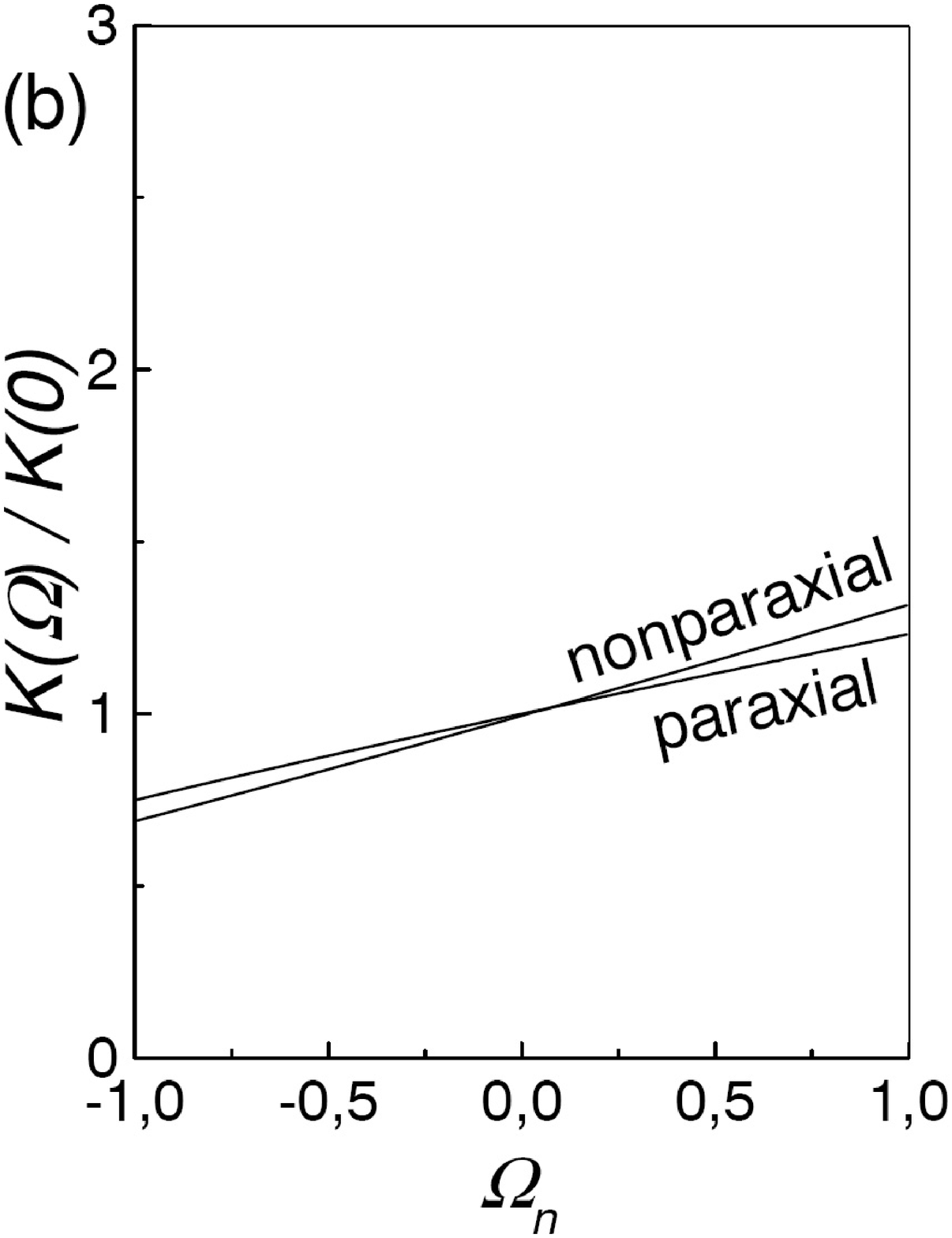}
\end{center}
\caption{\label{newfig3} (a) Paraxial and nonparaxial transversal dispersion curves of the modes of
carrier frequency $\omega_0=4$ fs$^-1$ with $\alpha=300$ mm$^{-1}$fs and $\beta=400$ mm$^{-1}$ in
fused silica ($k_0=19530$ mm$^{-1}$, $k'_0=4988$ mm$^{-1}$fs and $k_0^{\prime\prime}=77$
mm$^{-1}$fs$^{2}$). (b) The same as in (a) but only within the bandwidth of the shortest (widest
spectrum), single-cycle wave mode ($\Delta\Omega=\omega_0/2\pi$).}
\end{figure}

In a dispersive media, and under conditions (\ref{COND2}) and (\ref{COND3}) of quasi-luminality, we
can neglect in Eq. (\ref{NPDISP}) the terms $\beta^2$, $\alpha\beta$ and $\alpha^2$ in comparison
with $2k_0\beta$, $k_0\alpha$ and $2k'_0\alpha$, respectively, to obtain the approximate expression
\begin{equation}\label{DISPNP}
K(\Omega) \simeq \sqrt{2(k_0 + k'_0\Omega)\left(\beta + \alpha\Omega\right) +
k_0k_0^{\prime\prime}\Omega^2}
\end{equation}
for the nonparaxial dispersion relation of quasi-luminal modes. The first conclusion is then that
the paraxial dispersion curve [Eq. (\ref{DISP1})] may significantly differ from the nonparaxial one
[Eq. (\ref{DISPNP})], even if conditions (\ref{COND2}) and (\ref{COND3}) are satisfied. In fact, it
is not difficult to find set of parameters for which the nonparaxial dispersion curve is, for
instance, a vertical hyperbola, whereas the paraxial dispersion curve is an horizontal hyperbola
[see Fig. \ref{newfig3}(a)].

From a physical point of view, however, it is only the portion of the dispersion curve within the
mode bandwidth that is of relevance for the spatiotemporal mode structure, and, as our second
conclusion, this portion is approximately the same in the paraxial and nonparaxial approaches if the
additional condition (\ref{COND1}) of quasi-monochromaticity is also satisfied: Writing, for
transparent dispersive materials, $k_0/k_0'\approx \omega_0$, we obtain
\begin{equation}
K(\Omega)\approx \sqrt{2k_0\left(1+\Omega/\omega_0\right)\left(\beta +\alpha\Omega\right)
+k_0k_0^{\prime\prime}\Omega^2} ,
\end{equation}
or, in terms of the mode characteristic lengths,
\begin{equation}
K(\Omega)\approx\sqrt{2k_0\left[\left(1\!+\!\frac{\Delta\Omega}{\omega_0}\Omega_n\right)\left(L_p^{-1}\!+\!
L_w^{-1}\Omega_n\right) \!+\! \frac{1}{2}L_d^{-1}\Omega_n^2\right]} .
\end{equation}
Since $(\Delta\Omega/\omega_0)|\Omega_n|\ll 1$, the nonparaxial dispersion curve within the
bandwidth can be approached by the paraxial one, that is, by Eq. (\ref{DISP2}), as illustrated in
Fig.\ref{newfig3}(b) for the extreme case (widest possible bandwidth) of a single-cycle mode
($\Delta\Omega/\omega_0=1/2\pi$). In particular, we can affirm that the description performed in
Section \ref{CLASS} of quasi-monochromatic, quasi-luminal modes in terms of their characteristic
lengths is independent of the approach used.

\begin{figure}
\begin{center}
\includegraphics[angle=90,width=4.2cm]{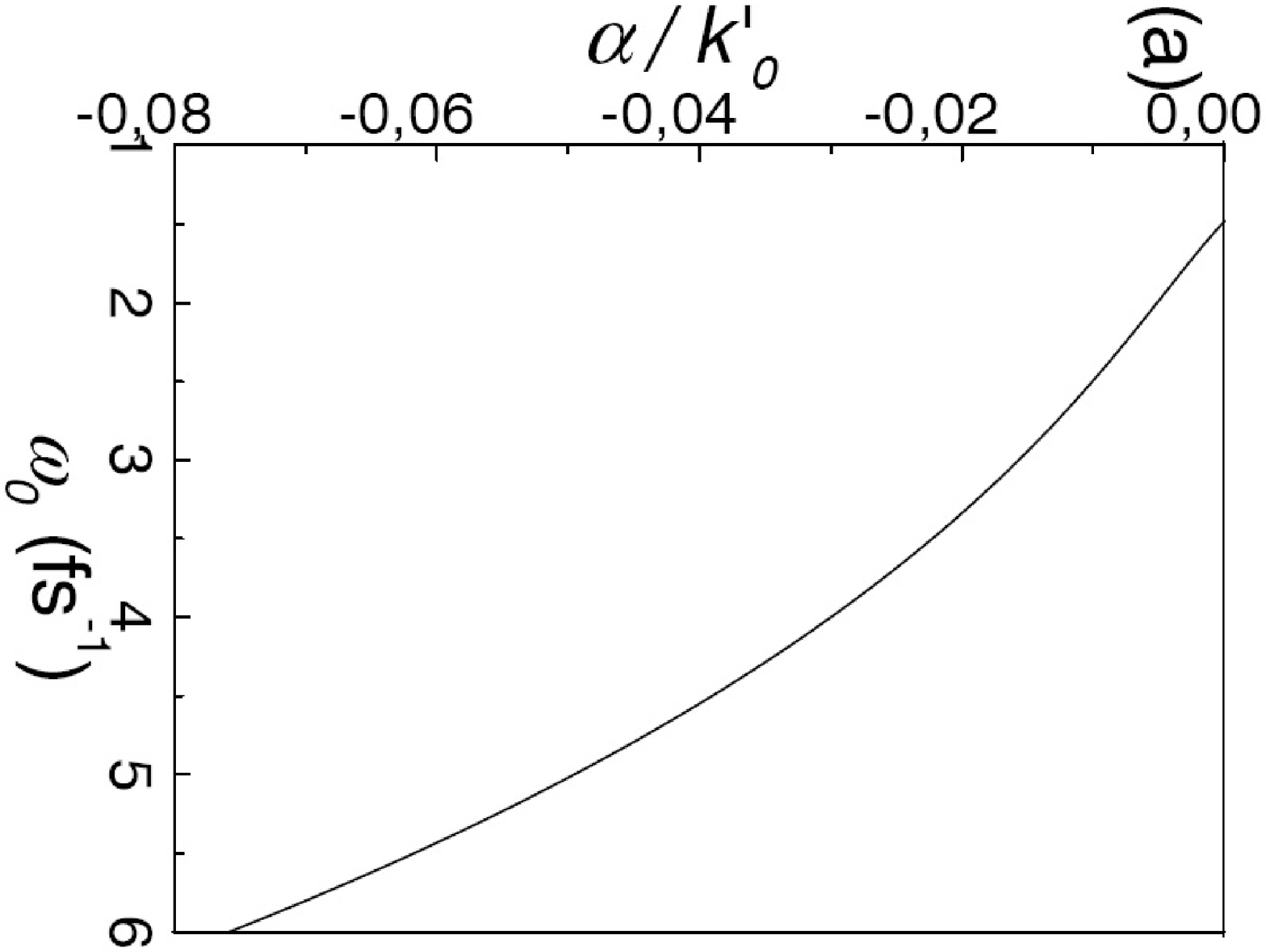}
\includegraphics[angle=90,width=4.2cm]{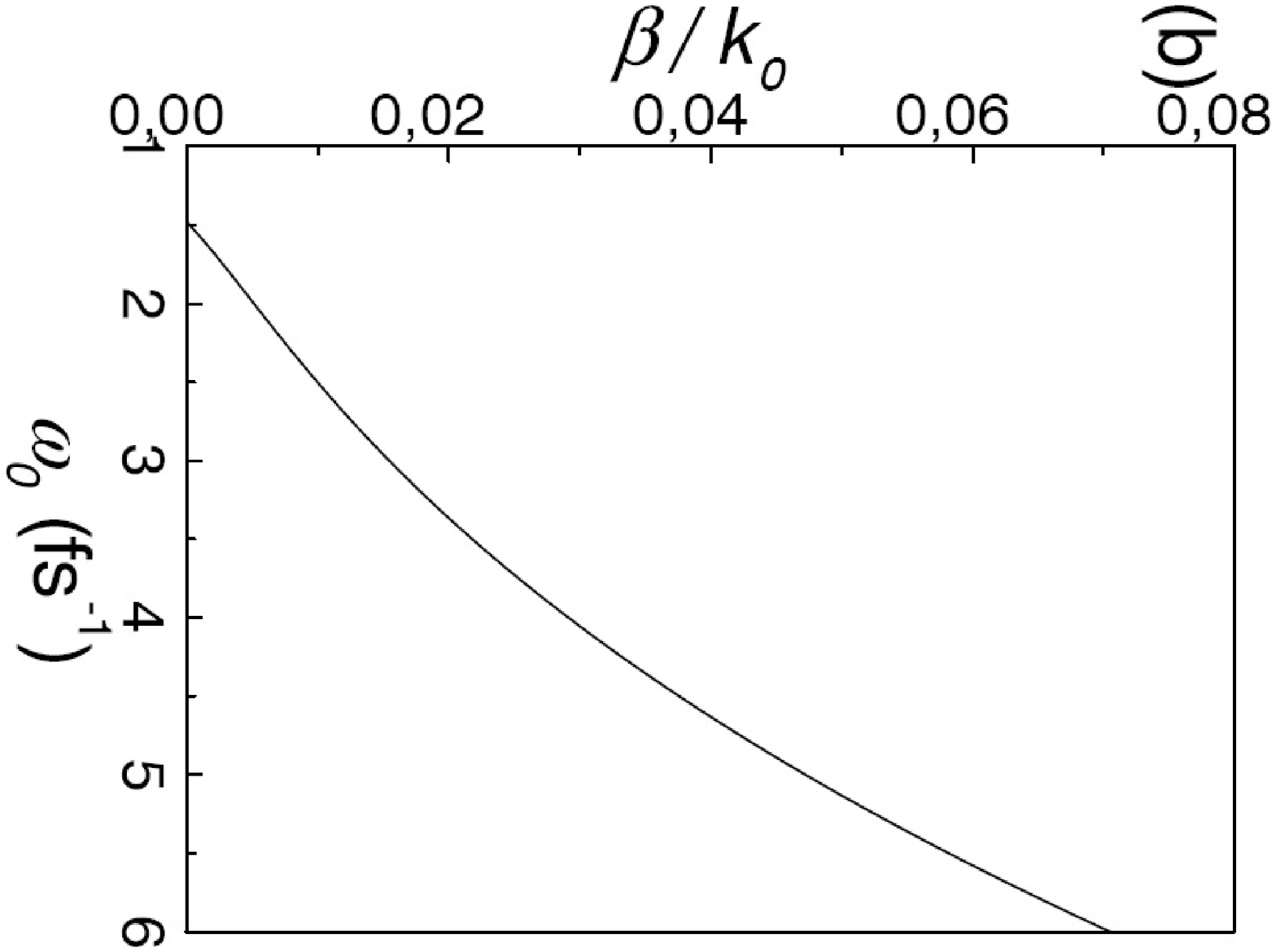}

\includegraphics[angle=90,width=4.2cm]{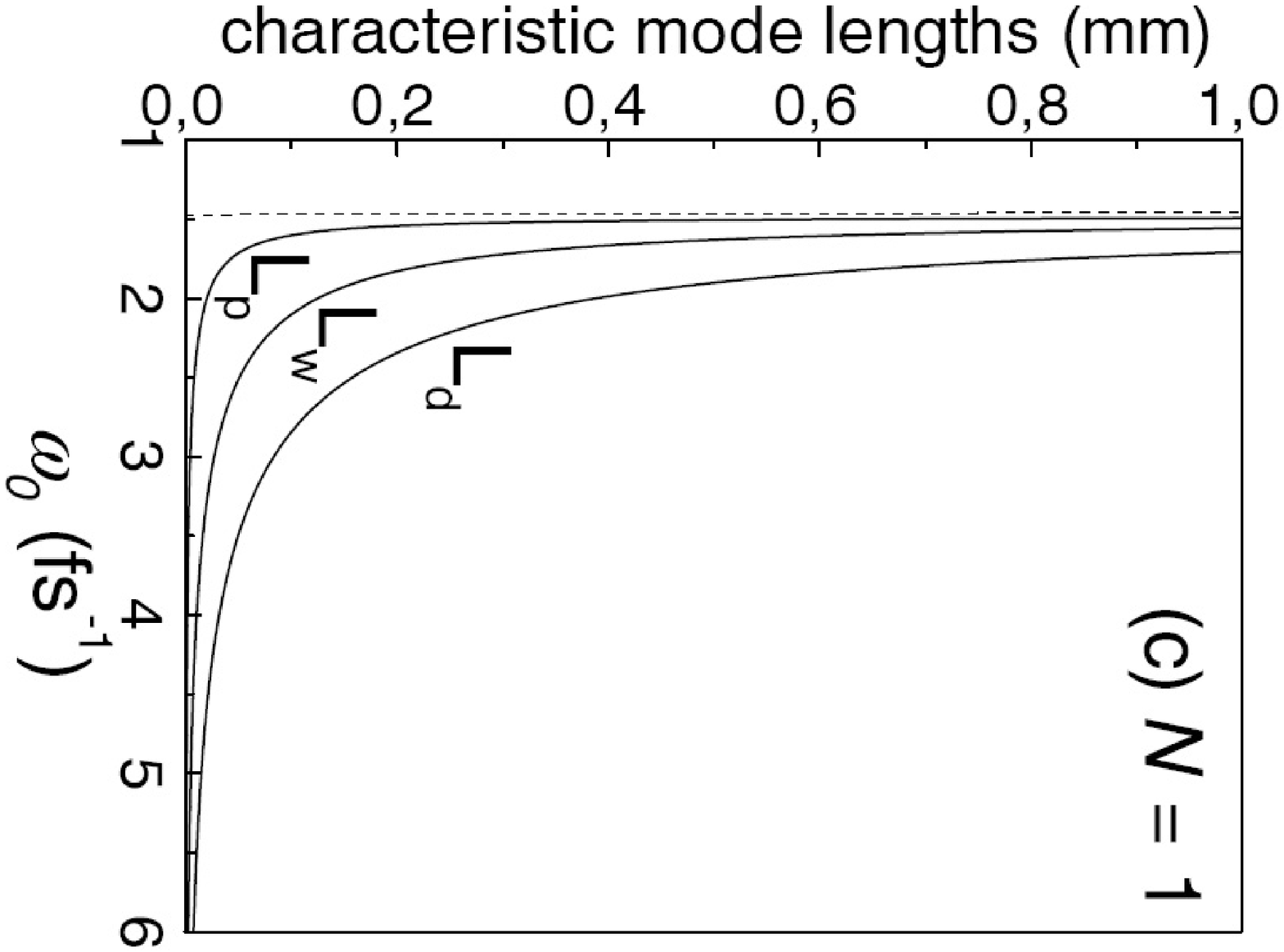}
\includegraphics[width=4.2cm]{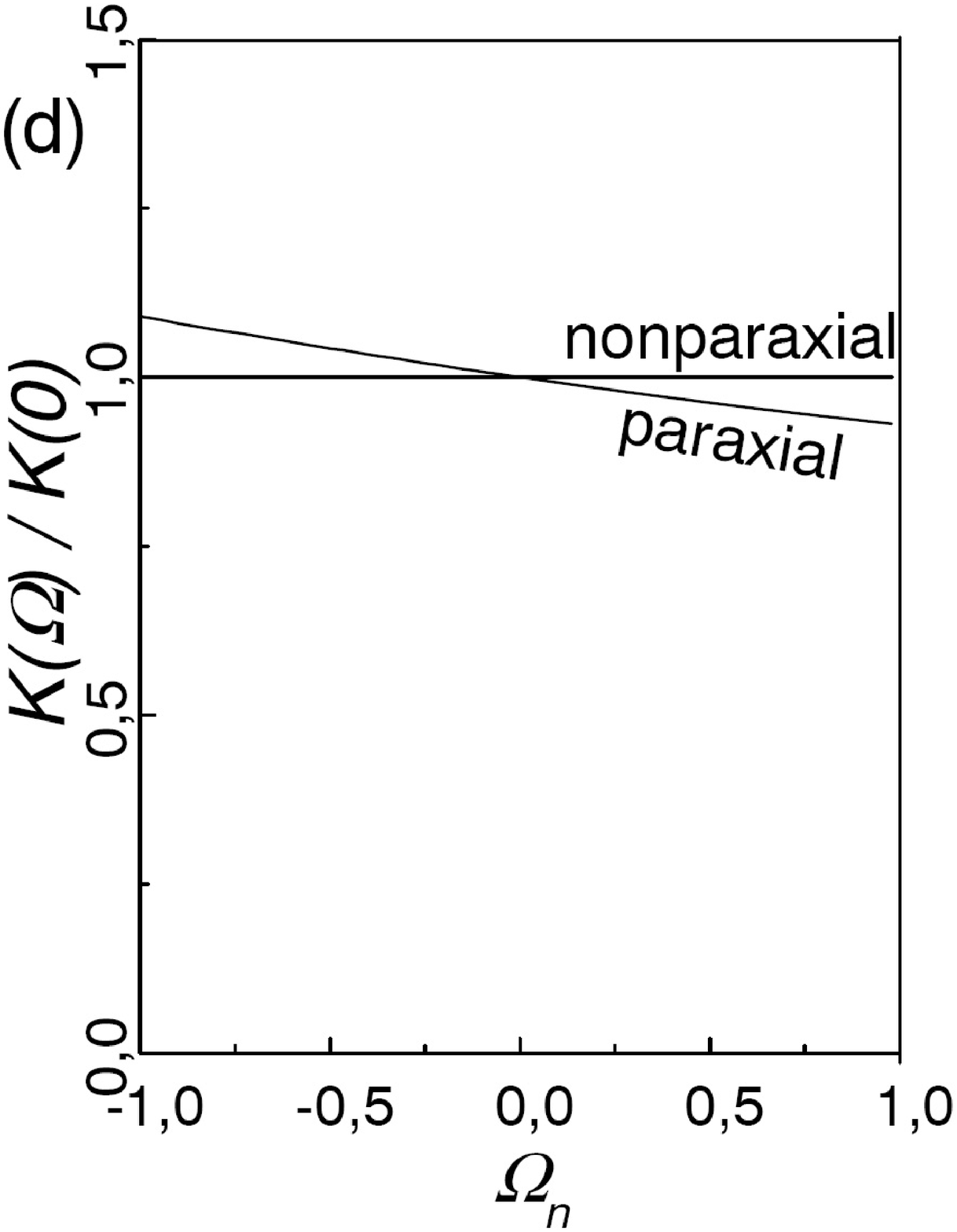}
\end{center}
\caption{\label{newfig4} (a) and (b) Values of $\alpha$ and $\beta$ from Eqs. (\ref{APBB}) and
(\ref{BPBB}) at different carrier frequencies in fused silica, with the refraction index obtained
from Ref. \onlinecite{HANDBOOK}. (c) Characteristic lengths for the limiting case of single-cycle
modes ($\Delta\Omega=\omega_0/2\pi$), with $\alpha$ and $\beta$ given by Eqs. (\ref{APBB}) and
(\ref{BPBB}) at different frequencies in fused silica. (d) For $\omega_0=2$ fs$^{-1}$ and
$\Delta\Omega=\omega_0/2\pi$, comparison between the paraxial and nonparaxial dispersion curves
within the bandwidth, given, respectively, by Eqs. (\ref{DISP1}) and (\ref{NPDISP}).}
\end{figure}

To illustrate the relationship between the paraxial and nonparaxial approaches, and the type of
results we can expect from the paraxial one, we consider wave modes of any bandwidth $\Delta\Omega$
propagating in normally dispersive media ($k_0^{\prime\prime}>0$) with
\begin{eqnarray}
\alpha &=& k'_0 - \sqrt{k_0^{\prime 2}+k_0k_0^{\prime\prime}}\simeq - k_0 k_0^{\prime\prime}/2k'_0 ,
\label{APBB}\\
\beta &=& - \frac{k_0\left(k'_0-\sqrt{k_0^{\prime
2}+k_0k_0^{\prime\prime}}\right)}{\sqrt{k_0^{\prime 2}+k_0k_0^{\prime\prime}}} \simeq k_0^2
k_0^{\prime\prime}/2k_0^{\prime 2} ,   \label{BPBB}
\end{eqnarray}
[see Figs. \ref{newfig4}(a) and (b) for propagation in fused silica], so that the nonparaxial
dispersion curve is, from Eq. (\ref{DISPNP}), the (exactly) horizontal straight line
\begin{equation}
K(\Omega) = K \equiv \sqrt{\frac{k_0^3k_0^{\prime\prime}}{k_0^{\prime 2}+k_0k_0^{\prime\prime}}}
\simeq \sqrt{\frac{k_0^3 k_0^{\prime\prime}}{k_0^{\prime 2}}} , \label{KPBB}
\end{equation}
and the corresponding nonparaxial wave modes are the dispersion-free, diffraction-free PBBs
$\Phi_{\alpha,\beta}(r,\tau+\alpha z) = f(\tau+\alpha z) J_0(K r)$ studied in Ref. \cite{PO01OL}.
The approximate equalities in Eqs. (\ref{APBB}), (\ref{BPBB}) and (\ref{KPBB}) hold for weakly
dispersive materials such that $k_0^{\prime\prime}\ll k_0^{\prime 2}/k_0$, in which case $\alpha$
and $\beta$ satisfy conditions (\ref{COND2}) and (\ref{COND3}) of quasi-luminality for the group and
phase velocities. As seen in Figs. \ref{newfig4}(a) and (b), this is the case of fused silica at any
visible carrier frequency.

For these PBBs, it is easy to see that the paraxial and nonparaxial descriptions become
undistinguishable, in spite of the apparent drawback that PBBs are no longer exact solutions of the
paraxial wave equation in dispersive media [when $k_0^{\prime\prime}\neq 0$, the paraxial dispersion
curve (\ref{DISP1}) is never an horizontal straight line]. In fact, when $k_0^{\prime\prime}\ll
k_0^{\prime 2}/k_0$, the relationship $|L_p|\ll|L_w|\ll|L_d|$ is satisfied for any mode bandwidth
down to the single-cycle limit [see Fig. \ref{newfig4}(c) for the case of fused silica].
Accordingly, these modes are of PBB type, that is, the paraxial dispersion curve within the
bandwidth can be approached by an horizontal straight line [see Fig. \ref{newfig4}(d) for
$\omega_0=2$ fs$^{-1}$ in fused silica]. Finally, the paraxial prototype PBB for these modes is
given, from Eq. (\ref{PBB}), by $\Phi_{\alpha,\beta}(r,\tau+\alpha z) = f(\tau+\alpha z) J_0(K r)$,
with $K=\sqrt{k_0^3 k_0^{\prime\prime}/k_0^{\prime 2}}$, that is, by the same expression as in the
nonparaxial approach.

\section{Conclusions}

Summarizing, we have described and classified the pulsed versions of Bessel beams with the property
of being localized and remaining stationary (diffraction-free and dispersion-free) during
propagation in a dispersive material with slightly super- or subluminal phase and group velocities.
As for the wave mode description, we have found the analysis of the transversal dispersion curve
$K(\Omega)$ to be an useful tool to understand the spatiotemporal mode structure. Wave modes have
been classified into three broad categories: PBB-like, eFWM-like,  and eX-like (eO-like) modes,
depending on the relative strength of their phase and group velocity mismatch with respect to a
plane pulse, and defeated GVD, as measured by the mode phase-mismatch length $L_p$, group-mismatch
length $L_w$ and the dispersion length $L_d$.

We have verified that the paraxial description leads to the same description and classification as
would be obtained from the more accurate nonparaxial approach when the conditions of narrow
bandwidth (\ref{COND1}) and of quasi-luminality (\ref{COND2},\ref{COND3}) are satisfied. All
previously reported optical Bessel beams, X waves, Bessel-X waves, or focus wave modes generated by
linear or nonlinear means satisfy indeed these requirements.

\section{acknowledgements}

The authors thank G. Valiulis helpful discussions, and acknowledge financial support from MIUR under
project Nos. COFIN 01 and FIRB 01.


\begin{thebibliography}{}
\bibitem{TRAPPRL2003} P. Di Trapani, G. Valiulis, A. Piskarskas, O. Jedrkiewicz, J. Trull, C. Conti
and S. Trillo, Phys. Rev. Lett. {\bf 91,} 093904 (2003); see also Phys. Rev. Focus, 4 September 2003
(http://focus.aps.org/story/v12/st7).
\bibitem{JEDRPRE2003} O. Jedrkiewicz, J. Trull, G. Valiulis, A. Piskarskas, C. Conti, S. Trillo and
P. Di Trapani, Phys. Rev. E {\bf 68,} 026610 (2003).
\bibitem{VA2001} G. Valiulis, J. Kilius, O. Jedrkiewicz, A. Bramati, S. Minardi, C. Conti, S. Trillo,
               A. Piskarskas and P. Di Trapani, in {\em OSA Trends in Optics and Photonics (TOPS),} QELS 2001
               TechnicalDigest, Vol.  57, (Optical Society of America, Washington D.C., 2001).
\bibitem{CONTIPRL2003} C. Conti, S. Trillo, P. Di Trapani, G. Valiulis, O. Jedrkiewicz, J. Trull,
Phys. Rev. Lett. {\bf 90,} 170406 (2003).
\bibitem{TRILLO} C. Conti and S. Trillo, Opt. Lett. {\bf 28,} 1251 (2003).
\bibitem{CONTI} C. Conti, Phys. Rev. E {\bf 68,} 016606 (2003).
\bibitem{DURNIN} J. Durnin, J.J. Miceli and J.H. Eberly, Phys. Rev. Lett. {\bf 58,} 1499
(1987).
\bibitem{SO96} H. Sonajalg and P. Saari, Opt. Lett. {\bf 21,} 1162 (1996).
\bibitem{SO97} H. Sonajalg, M. Ratsep and P. Saari, Opt. Lett. {\bf 22,} 310 (1997).
\bibitem{PO01OL} M. A. Porras, Opt. Lett. {\bf 26,} 1364 (2001).
\bibitem{PO02OC} M. A. Porras, R. Borghi, M. Santarsiero, Opt. Commun. {\bf 206,} 235 (2002).
\bibitem{ORLOV1} S. Orlov, A. Piskarskas, A. Stabinis, Opt. Lett. {\bf 27}, 2167 (2002);
S. Orlov, A. Piskarskas, A. Stabinis, Opt. Lett. {\bf 27}, 2103 (2002).
\bibitem{POOL2003} M. A. Porras, S. Trillo and C. Conti, Opt. Lett. {\bf 28,} 1090 (2003).
\bibitem{POPRE2003} M. A. Porras, G. Valiulis and P. Di Trapani, Phys. Rev. E {\bf 68,} 0166 (2003).
\bibitem{LU} J. Lu and J.F. Greenleaf, IEEE Trans. Ultrason. Ferroelectr. Freq. Control
{\bf 37,} 438 (1990).
\bibitem{SAALP} P. Saari and H. Sonajalg, Laser Phys. {\bf 7,} 32 (1997).
\bibitem{SAAPRL} P. Saari and K. Reivelt, Phys. Rev. Lett. {\bf 79,} 4135 (1997).
\bibitem{VALIULIS} G. Valiulis, Optical Nonlinear Process Research Unit, internal report (unpublished).
\bibitem{HANDBOOK} See for instance, {\em Handbook of Optics,} Vol. II (second Edition) McGraw-Hill (New York,
1995).
\bibitem{GRA} I. S. Gradshteyn and I. M. Ryzhik, {\em Table of integrals, series
and products},  Academic, (NY, 1965).
\bibitem{BRI} Brittingham, J. Appl. Phys. {\bf 54,} 1179 (1983).
\bibitem{ZIOL} R.W. Ziolkowski, Phys. Rev. A {\bf 44,} 3941 (1991).
\bibitem{SAAJOSAA} K. Reivelt and P. Saari, J. Opt. Soc. Am. A, {\bf 17,} 1785 (2000); Phys. Rev. E
{\bf 65,} 046622 (2002).








\end{thebibliography}
\end{document}